\documentclass[utf8]{article}
\usepackage[a4paper, total={6in, 8in}]{geometry}
\usepackage{url,hyperref,lineno,microtype,subcaption,natbib}

\begin{document}
\title{An observer's view on the future of asteroseismology}
\author{Margit Papar\'o\thanks{e-mail: paparo@konkoly.hu}\\
{\normalsize Konkoly Observatory, MTA CSFK,
        Konkoly Thege M, u. 15-17., H--1121 Budapest, Hungary}}
\maketitle

\begin{abstract}

Scientific research is a continuous process, and the 
speed of future progress can be estimated by the pace of finding 
explanations for previous research questions. In this observer’s based 
view of stellar pulsation and asteroseismology, we start with the 
earliest observations of variable stars and the techniques used to 
observe them. The earliest variable stars were large amplitude, radial 
pulsators but were followed by other classes of pulsating stars. As 
the field matured, we outline some cornerstones of research into 
pulsating star research with an emphasis on changes in observational 
techniques. Improvements from photographs, to photometry, CCDs, and 
	space telescopes allowed researchers to separate out pulsating stars from 
other stars with light variations, recognize radial and nonradial 
pulsation courtesy of increased measurement precision, and then use 
nonradial pulsations to look inside the stars, which cannot be done 
any other way. We follow several highlighted problems to show that 
even with excellent space data, there still may not be quick 
theoretical explanations. As the result of technical changes, the 
structure of international organizations devoted to pulsating stars 
has changed, and an increasing number of conferences specialized to 
space missions or themes are held. Although there are still many 
unsolved problems, such as mode identification in non-asymptotic 
pulsating stars, the large amount of data with unprecedented precision 
provided by space missions (MOST, CoRoT, Kepler) and upcoming missions 
allow us to use asteroseismology to its full potential. However, the 
enormous flow of data will require new techniques to extract the 
science before the next missions. The future of asteroseismology will 
be successful if we learn from the past and improve with improved 
techniques, space missions, and a properly educated new generation.



\hspace{1cm}

\end{abstract}

\section{Introduction}

Let me start with ancient times to show how slow was the
progress over centuries without serious technical devices.
The stars of the night sky, especially the brighter ones, have always been part 
of the life of people. Initially, the most prominent objects whose celestial 
movement was perceptible were planets that were considered to be gods who had a direct impact on 
their lives. The fixed stars were formed into constellations 
that depicted mythical figures. Over the centuries, however, individual 
observations had accumulated and founded the science of astronomy. Although 
stars still occupied a special place in the lives of people, the Sun and the
distant stars  were no longer considered gods and myths, but simple cosmic 
objects. With these few sentences, I have described centuries, first, those years 
when stars were only seen with naked eyes, and those when telescopes were already
built, but observations were recorded only by drawings, such as 
spots on the Sun or phases of the planets. Then the development and 
appearance of the photographic technique
in the observation of stars allowed us to capture their current status, how 
bright or dim they are. Finally, the research of variable stars began.

In this article I would like to show how the separation of stars have developed into independent groups as a result 
of technical progress (increasing precision) and how the actual technique (photographic, photoelectric or CCD) influenced which problems (group of stars or individual stars connected to the
field of view of the actual instrument) were 
examined. I also present how scientific results (not properly resolved and biased frequencies) and requirements (continuous, long observations compared to the pulsation periods) have led to the 
need for spacecraft.
I do not intend to provide a detailed and complete overview of the scientific 
results of the last 60-70 years, but I want to highlight  
the dominant directions of research in a given decade up to the space era. 
The results
and raised questions of a given decade turned out to be the base of the research lines of the next
decade. From decade to decade, unresolved issues have shown what
the hot lines of scientific research in the era of space telescopes will be and what the future of astroseismology holds. I do not intend to give a complete theoretical overview, I only want to emphasize and show that there can be no substantial progress without theoretical interpretation.

\section{Pre-history of asteroseismology}

In the last centuries naturally the largest light variations of the brightest stars were discovered first.
Mira ($o$ Ceti) was the first to be discovered, with observations dating back to the 1600's \footnote{https://www.aavso.org}. 
A detailed historical summary of the observation of Mira was given by \citet{Hoffleit1997} for the
400 year anniversary of the discovery.
The light variability of $\delta$ Cep was discovered in the 18th century \citep{Goodrick1786}.
The variability of RR Lyrae, along with 64 additional newly discovered variable stars, was 
announced 40 years later \citep{Pickering1901}. The radial velocity
variation of $\beta$ Cephei was noticed by \citet{Frost1906}.
The date of the first discovery of a $\delta$ Scuti type variation is more 
difficult to determine. The change of $\delta$ Scuti itself was announced by 
\citet{Fath1935}, and radial 
velocity measurements were also published in that year \citep{Colacevich1935}. 
However, $\delta$ Scuti type stars as a new group were identified only after
another twenty years \citep{Eggen1956}. 
The process is clear, we are talking about a new group if we have found more 
stars with a similar variation. The criteria include 
the degree of variation and the star's location on Hertzsprung-Russell diagram. In the subsequent years 
and decades several new groups of variable stars were discovered and 
identified. I will follow how the knowledge of the groups improved from decade
to decade at first for the early recognized groups and later for the newly
identified groups, too.

Of course, the light variability of stars aroused the interest of 
theoretical physicists. Many of Eddington's works dealt with the 
theory of pulsation, which he summarized in a book \citep{Eddington1926}.
He predicted that the light variability generated by the 
pulsation of the stars would provide information about the inner structure 
that otherwise we cannot acquire. Asteroseismology is exactly the method in which 
the inversion of the star's light variability, getting at first their 
frequency content, leads to the physical parameters inside stars, i.e.
temperature, pressure, density, speed of sound and chemical composition along the 
radius. The method, called helioseismology, has worked successfully in the case of
the Sun. On the other hand distant stars do not easily share their secrets with us.
We have determined the frequency content of many stars and their global physical 
parameters, but inversion has not been achieved so far for most kinds of pulsating stars, so the dependence of the physical 
parameters on the radius of the star has not yet been determined, except for the Sun and compact stars.

In the following I show how we have come closer to our goal from decade to decade, and how much 
the technical development of the given age and the related accuracy of the
measurement influenced what problems were investigated by previous researchers.
I am convinced that knowing the work of predecessors can have a profitable 
impact on today's research. This article is not a complete review
of each field of pulsating stars that would require a more regular structure of the paper. I decided to use my life as a guideline when I was invited to present my view on the future of asteroseismology (another personal review). I decided to begin the summary at the epoch when
I was born and the science field started to attract more attention.

\section{Results between 1950-1960}

From the point of view of technical development, the middle decade of the 20th 
century was particularly impressive and productive. The {\bf photographic technique} 
was still used; moreover, \citet{Cuffey1954} designed and built a new photometer at the 
beginning of this decade. One of the two beams came directly 
from a control lamp while the other led to an oscilloscope via the iris and the photographic plate. Matching the two rays provided the star's brightness.
The photometer produced 0.003-0.13 mag accuracy on the Mt Wilson 100-inch 
telescope, depending on the brightness of the star and its position on the 
plate. The photometer proved to be so successful that I started working with 
such type of photometer in 1974 when I began my scientific career at  
Konkoly Observatory, Hungary. Of course, the accuracy was worse due to the use 
of a smaller telescope, but the error of the finally accepted physical 
parameters were seemingly reduced by averaging many plates obtained 
on the same field.
The enormous advantage of the photographic photometry was the large area 
covered with wide-angle telescopes. The most well-known project was the
all-sky survey of the 48-inch Palomar Schmidt camera resulting in the Palomar 
catalog. Of course, this technique was most advantageous for the discovery of 
stellar variability and for the investigation of stellar clusters as well as galaxies.

Naturally, there was the need to achieve greater precision for individual stars, 
which called for a new technical solution. Photoelectric photometers started to be 
produced in the leading astronomical observatories/institutes.
A photoelectric photometer generally consisted of an aperture in the focal
plane of the telescope to isolate the light from a single star. The light then
(usually) passed through a filter of known passband to a photomultiplier which
converted photons to pulses of electrons which could be measured as a weak
current or, better, counted as a series of pulses. Photometry using such device
consisted of pointing the telescope alternatively at a variable star, a 
comparison star and piece of sky without visible stars (to measure the sky
background).
This inevitably put a heavy burden on the observer when he/she had to work as 
quickly as possible, as long as possible on many nights one after another
to answer certain questions of the star which was being investigated.
It was \citet{Walraven1952} who designed a photoelectric photometer which 
removed the human efforts by rendering the actions otherwise 
performed by the observer automatic. It was attached to the 16-inch Rockefeller 
telescope of the Leiden Station of Johannesburg. However,
it took some time for photoelectric photometers to appear in the
observatories all over the world. Scientists of communist countries, for 
example, suffered restrictions on high-tech electronic devices due to political concerns.
I myself started to work with a normal photometer in 1980 at Konkoly
Observatory (moving the telescope by hand from variable, to comparison and 
to background), when I changed my interest from photographic to photoelectric
photometry, since I was delighted by the higher precision. The photoelectric  
photometer was built in our observatory by Géza Virághalmy using a photomultiplier tube 
which was personally donated, without any official permission, to 
the director of Konkoly Observatory, to László Detre by Th. Walraven when he 
visited Hungary.
Both photographic and 
photoelectric photometry used filters in different wavelength to get more 
information on the stars in general, and on the characteristics of the
pulsation. The two basic photometric systems, Johnson UBV \citep{Johnson1955} 
and Strömgren ubvy \citep{Stromgren1954} were established during this decade. 

The development of {\bf spectroscopy} had gone through a similar process. In the past,
only an objective prism was employed for the simultaneous photography of the 
spectra of large numbers of faint stars. After World War II, \citet{Wood1946}
reported that it was possible to construct transmission gratings
which throw as much as ninety percent of the light into a single spectrum.
In the next years, the technical development accelerated. In ten years
\citet{Harrison1955} reported that echelle 
spectrographs had been produced in six forms. Some years later an electronic 
camera was developed by Lallemand and Duchesne and was installed at the focus
of Coudé spectrograph of the 120-inch reflector at Lick Observatory \citep{Lallemand1960}.
The sensitivity of the photocathode was 10-20 \% instead of 0.1 \% of the 
photographic emulsion, which meant a gain in speed by a factor of about 100.
The 2-dimensional classification of stellar spectra, the Morgan-Keenan system, was also
established in this decade \citep{Keenan1951, Johnson1953}.

The higher precision of the photoelectric technique and the higher sensitivity
of spectroscopy led to the discovery of 
more and more variable stars of lower amplitude and shorter period.
Groups of similar stars were recognized, such as the 
Cepheids, RR Lyrae, cluster variables, dwarf cepheids, short period 
$\delta$ Scuti, $\beta$ CMa 
and $\beta$ Cephei variables. There was no clear-cut separation between the groups, but the
nomenclature changed when more information was collected on the groups. The
research field was in the data collection phase. On the Harvard patrol plates
\citet{Payne1954} discovered 51 RR Lyrae stars, 135 classical Cepheids 
and nine variables of a similar period that belonged to Population II.
The two groups of stars, Pop I and Pop II were recognized
by \citet{Baade1944}. Pop I stars are younger and are located in the spiral 
arms, while the Pop II stars are older and are located in globular clusters.
Later, different metallicities were found for the groups. 
In the M3 globular cluster the RR Lyrae stars and those non-variable stars
were measured which border the variable ones on the HR diagram
\citep{Roberts1954}. The resulting color-magnitude diagram showed
that all the cluster-type variables fell in a distinct region which was not
occupied by non-variable stars. So, there was a sharp separation 
between variable stars and non-variable stars. The basic question whether there
are non-variable stars in the instability strip was raised
for the first time, and has been investigated up to now.
If the answer is yes, then what is
the physical reason that some stars with similar physical parameters 
(temperature and surface gravity) show pulsation, while the others do not?
This is one of the issues that we can follow from decade to decade to see how fast we can provide an answer to a question, if we can manage it at all.

All the available technical possibilities were used in the {\bf data collection} for
RR Lyrae stars during this decade.
For the end of the decade photoelectric
observations in {\bf V and B} were collected for nine RR Lyrae stars 
\citep{Spinrad1959}. \citet{Preston1959} got spectra of low dispersion (430 {\AA}/mm at 
$H_\gamma$) for hundreds of RR Lyrae stars. He derived the $\Delta$s index to 
characterize the metal content of the stars at minimum light. This way a  new possibility
was opened for comparing the stars according to their evolutionary status.
Additionally, the discovery and analysis of 49 Cepheids in the Small Magellanic Cloud \citep{Arp1958}
opened the new question of whether, at a given period, the Cepheids in the
galaxy and SMC had the same luminosity, mass or chemical composition.
Period-luminosity (P-L) relation for Cepheid variables had been exclusively
used to find distances to nearby galaxies, since its discovery by Miss Leavitt
\citep{Leavitt1912}. However, the precise photometry of \citet{Arp1955} 
pointed out that the
scatter in the P-L relation was real, and not due to the photometric uncertainty.
According to \citet{Sandage1958}, there was good evidence that Cepheids in the globular
cluster obeyed a mean P-L relation, which was at least 1.5 mag fainter than the P-L
relation for classical Cepheids. Both types of population and pulsation mode seemed
to have an influence on the P-L relation.
Although the Population I and II Cepheids were identified 
\citep{Payne1954}, the type of pulsation modes was not clearly 
determined. 

For other type of variables less information was obtained, which, however, were compared to each other.
In the early years of this decade, before the identification of the group of $\delta$
Scuti stars \citep{Eggen1956}, $\delta$ Scuti itself was compared to classical
Cepheids, to cluster-type variables, and to $\beta$ Canis Majoris stars. Similarities or differences were searched for.
The maximum brightness of $\delta$ Scuti occurred very close to the time of minimum radial
velocity \citep{Struve1953a}, as in the classical Cepheids and in cluster-type
variables. A range of about 0.005 magnitude in color was found, and the star
proved to be bluer at maximum light, as were the Cepheids and certain 
$\beta$ CMa stars. By the end of the decade seven members of the $\delta$
Scuti group were confirmed. Both fundamental and first overtone modes
were reported observationally \citep{Fitch1959}. 
The $\beta$ CMa and $\beta$ Cephei stars were compared to determine whether they belonged to
the same group or not \citep{Struve1953b}. The possible members displayed
similarities and differences concerning the number of periods, the
variation in the line profile, the difference in luminosity and sudden
change in the main period. Definitely more data were needed to determine
the real nature of these types.
However, the amount of knowledge collected for Cepheids seemed to be so convincing that
in 1954 the Irish
Astronomical Journal presented news based on the
paper of \citet{Grand1954} that "the theory of 
internal structure and pulsation of Cepheids as well as the structure of
their atmospheres, is now based on solid ground".
The statement was absolutely truthful in that decade, but we will see how much
new information was collected for Cepheids in the next decades.  This case teaches us that our current knowledge serves only as the base for the next improvements, although we might feel at each epoch that we have found the truth.

\section{Results between 1960-1970}

There was no dominant technical development in 1960's. Everything was
available, both photographic and photoelectric photometry as well as spectroscopy
for collecting more information on pulsating stars. The photographic technique allowed us to investigate larger structures of stars (clusters and galaxies) with less accuracy, while the photoelectric technique was used to get measurements of higher precision for individual stars. I emphasize here the importance of the interaction of observation and theoretical interpretation. The observations collected
in the previous decade inspired much {\bf theoretical work}. According to 
\citet{King1968} the theoretical study of pulsating variable stars had
been considerably advanced in the previous ten years. The progress was attributed
in large part to the availability of modern electronic computers, which
made it possible to treat the equations appropriate to radial pulsations in
their fully nonlinear form. They emphasized that the most important and 
far-reaching result of Eddington was
the well-known period-mean density relation, P($\rho$/$\rho_{\odot})^{1/2}$=Q, where
Q was a constant. \citet{Zhevakin1963} was the first who found that the interaction
of the He II ionization with pulsation led to a sufficiently strong unstabilizing
force that could overcome damping and drive pulsation. The evolutionary
path to the main sequence was calculated by \citet{Iben1965a}, and later in many papers
he presented the evolutionary path from the main sequence through core helium
burning for stars of different masses \citep{Iben1965b}. As a major result,
a theoretical
period-luminosity relation was derived by \citet{Hofmeister1967} in an evolutionary
calculation of Cepheids.
Compared with the observational data, a good
match was found for the fundamental mode. The period-luminosity relation
has been used more carefully since then.
Stars at a different evolutionary stage served as good distance
indicators to the spiral arms and the younger parts of the Galaxy (Pop I
classical Cepheids), while the RR Lyrae and Pop II Cepheids (W Virginis) stars that range 
from intermediate to old age provided an important clue to the kinematics 
and chemical history of the Galaxy. \citet{Wesselink1963} raised a 
quite relevant question whether 
Pop I RR Lyrae variables exist. There
was no answer for this question at that time. Although, Pop I RR Lyrae stars
were not found in the next decades, the field RR Lyrae stars of higher metallicity were discovered.
    Combining the period-density law and the expressions for density and 
luminosity, the pulsation constant (Q) could be calculated from 
photometric observables (absolute magnitude, effective temperature, 
surface gravity and the pulsation period) \citep{Danziger1967}. 
The theoretical model of \citet{Christy1966} for RR Lyrae
resulted in the pulsation constant for the fundamental and
some overtone modes. The transition period ($P_{tr}$=0.057(L/L$\odot$)$^{0.6}$), where the fundamental pulsation
changed to the first overtone depended only on the luminosity and was independent of
mass and composition. Following this a model for classical Cepheids was produced by
\citet{Stobie1969}. Theoretical models were inevitable to
interpret the observational data for the different type of variables.
The pulsation constant and the period ratios became an 
important tool for determining the type of pulsation in pulsating stars.

Despite the great improvement in the theory, there 
remained questions: what is the limiting amplitude and what is the
mechanism behind it? Or what is the effect of convection on pulsation
stability? It seemed that a definite theory of time-dependent convective transfer
was missing.
In some types of pulsating stars even the cause of the exciting mechanism was
not known (Mira variables and other type of red variables). According to 
\citet{Christy1966}, there was no indication that the $\beta$ Cephei variables could be
accounted for by the mechanism responsible for RR Lyrae and Cepheid pulsation.
The role of the
non-radial oscillations of the stars was not explained. New observations were
urgently needed to step further in theoretical explanations. On the issues of the amplitude limitation and the convection also show how long a time span and how much observational effort is needed to solve the basic problem of pulsation.

Apparently, sufficient numbers of {\bf Cepheids} were known in the '60s, because there were only a 
few surveys initiated and mostly in extragalactic systems. \citet{Gaposchkin1962}
found 223 variables in the Andromeda Nebula. 59 percent of the 
variables turned out to be Cepheids, 28 percent were reddish and red
irregular variables and five percent eclipsing binaries. \citet{Mich1965} obtained
UBV photometry for eight Population II Cepheids. \citet{Kraft1967}
summarized the importance of the spectral lines in the atmospheres of
Cepheids, which were able to give information on the chemical composition and on
the velocity of ordered motion of the gases constituting the atmosphere. 
The physical problems were discussed
in connection with the atmosphere of a pulsating star. An obvious additional
effect was the broadening of the spectral lines due to pulsation. The 
reasons were not only the geometrical effect, and a phase dependent
velocity gradient, but also a possible change in the velocity parameters usually
associated with the phenomenon called 'turbulence'. 

There was such an urgent
need to produce time-dependent convection instead of the mixing-length
theory \citep{Bohm1958} that three serious attempts were published
in the 60's \citep{Gough1967, Unno1967, Castor1968}. (The treatment
of convection is definitely going to be a critical topic for asteroseismology 
in the space era.) It is connected to the recognized problem that the
red edge of the instability strip is determined by convection, which 
was noticed in the RR Lyrae model of \citet{Baker1965}.

For {\bf RR Lyrae stars} there were some surveys and all techniques were used 
during this decade. 62 RR Lyrae stars
were found on the southern hemisphere \citep{Clube1969}, and 77 new RR Lyrae stars
appeared in the Lick survey \citep{Kinman1966}. \citet{Sturch1966} obtained
UBV photoelectric photometry of more than one hundred stars to derive the color
indices at minimum light. \citet{Paczynski1966} used the Lallemand photomultiplier
for getting three color photometry of RR Lyrae stars, while \citet{Preston1965}
carried out simultaneous spectroscopic and photoelectric observations
for the RR Lyrae itself. \citet{Babcock1956} discovered a variable magnetic field
in the spectrum of RR Lyrae. 

The multi-periodicity of RR Lyrae stars had been known for years,
where the long periodicity was always of the order of 100 times longer than  
the fundamental period. The modulation phenomena of RR Lyrae stars was 
named as the {\bf Blazhko effect} in an ADS search only from 1959.  A
suggestion for possible cause of the long-period variability, the oblique
rotator model of an eruptive variable
with an oscillating field was presented by 
\citet{Balazs1962}. Despite the observations of long-period
variability in RR Lyrae stars, \citet{Christy1966} did not find a 41-day modulation
in his RR Lyrae model. Basically, he regarded the modulation phenomenon as a
superficial manifestation rather than a deep-seated deviation from 
his model. As we will see, trying to interpret the Blazhko modulation
in RR Lyrae stars extends even to the present day using space data. The Blazhko effect (discovered by \citet{Blazhko1907}) is a valuable indicator, too, to track the speed of solution in science.

The observed position of classical Cepheids, RR Lyrae stars, $\delta$ Scuti
stars and dwarf Cepheids on the HR diagram formed a continuous sequence in the log g versus
log T$_{eff}$ plane. It was not surprising that there was a confusion in
naming. Fortunately, the confusion was settled in this decade.
\citet{Wesselink1963} declared that Cepheids with periods shorter than a
day, which were often called 'cluster' variables, were more suitably named
after their prototype RR Lyrae. According to \citet{Bessel1969a}, the name of
dwarf Cepheids was misleading and naming after of the prototype, AI Velorum
was more appropriate. However, \citet{Breger1969a} suggested that it 
was reasonable to define the dwarf Cepheids as low-mass $\delta$ Scuti-like 
variable stars. Two subgroups of $\delta$ Scuti stars were suggested.
The stars in one group had a mass of 1.9 M$_\odot$ \citep{Bessel1969b},
while the mass of the stars in the other group was 0.5 M$_\odot$ \citep{Bessel1969a}.
However, the future will show that this was not the final grouping of
$\delta$ Scuti stars. Due to the overlap of the evolutionary paths from
the main sequence to the later evolutionary stages, this part of the HR diagram is rather complex.
To see the similarities or differences among the different type of variables, 
the relation between
the amplitudes of the light and radial velocity curves were compared to each
other \citep{Leung1970}. The slopes for the RR Lyrae and $\delta$ Scuti
were nearly the same, while the slope for the $\beta$ Cepheid was substantially
smaller. The light and radial velocity data were consistent with the
interpretation that the variability of $\delta$ Scuti stars arose
from radial pulsation similar to the RR Lyrae and the $\delta$ Cepheid stars.
The different groups of pulsating stars being so closely located warns us that pulsation is very common at certain ages and it is only us who separate them into groups (that is, put them into categories).
 
After the definition as a group in the 50's, the {\bf $\delta$ Scuti stars} were intensively investigated in this decade. 
\citet{Millis1967} obtained photoelectric photometry, while \citet{Danziger1967}
carried out spectrophotometry for new short-period variables. 
\citet{Breger1969a}
tested 213 bright field stars for short-period variability based on photoelectric measurements.
19 new variables were found. Narrow-band (Strömgren) photometry and spectroscopic data 
were collected for most of the 39 $\delta$ Scuti variables published until then. 
The exponential frequency-amplitude distribution predicted that most stars
should vary by a few thousandths of a magnitude with definite periods. 
The question also emerged in connection to $\delta$ Scuti stars whether all the stars in the instability strip might really be
variable or not. This question was raised in the 50's and this is still an open question in the space era. If we compare the time scale of this unsolved problem to the financial support of 3-5 years of a project, it is obvious that the solution of the basic problems need much more time than the duration of a project. 
It is even worse if the continuation of the same research projects are not supported in the next application.
The empirical
$\delta$ Scuti instability strip and the period-luminosity-color relation
were established \citep{Breger1969b} in this decade. Actually, only slightly more than 20 percent
of the stars showed regular variability larger than 0.010 magnitude. The physical
parameters were derived and investigated in connection to the pulsation.
In conclusion, the $\delta$ Scuti stars had
normal mass, but a wide range of metal abundances and rotational velocities. Metal
abundance and binarity appeared to have no influence on the incidence of
variability. Metallicity and rotation had no effect on the period.
The small-amplitude variables shared the high rotational
velocity of the non-variables, while the incidence of larger amplitudes depended
critically on low rotational velocity. The classical Am stars were stable
against the pulsation \citep{Breger1970}. There was a remarkable improvement 
in understanding the pulsation in this group in this decade.

The $\beta$ CMa and $\beta$ Cephei groups merged to a single group by
this decade.
24 new {\bf $\beta$ Cephei stars} were found in a search by \citet{Hill1967}. 
Altogether
there were 41 members of the group. The observations extended considerably the
limits of spectral type, luminosity class, absolute magnitude, periods
and rotational velocity. The $\beta$ Cephei stars occupied no preferred
position in relation to normal stars, i.e. many of the non-pulsating stars were at 
the same stage of evolution as the $\beta$ Cephei stars. However, the
driving mechanism of the excitation was still unknown.

Observationally it was evident that both $\delta$ Scuti and $\beta$ Cephei
variables showed long-period modulation \citep{Fitch1967} and beat phenomena \citep{Breger1969b}. 
Even the amplitude modulation of RR Lyrae stars was mentioned to be connected to this phenomenon.
The various interpretations
showed large variety: 
(1) a tidal deformation by a faint
component, (2) two nearly equal periods excited by rotation, (3) simultaneous
excitation of a large number of radial modes by resonant interactions, (4)
rotational coupling of a purely radial and a purely non-radial oscillation
modes, (5) oblique rotator, (6) the radial pulsation of a non-spherical star
which rotates and presents different aspects of its non-spherical appearance
during the course of the rotation period, and (7) pure non-radial pulsation 
\citep{Ledoux1951}. 
The task to answer the remaining
questions raised by the theory and to step further in understanding the stars
with 'light modulation' was obvious for the next decade.

\section{Results between 1970-1980}

Photoelectric photometers, beside the traditional photographic possibilities,
spread world-wide in the smaller observatories/institutes, 
not only in the major astrophysical centers. Using 
photoelectric photometers we gained higher precision, but we lost the wide 
field of view of the photographic plates. The photoelectric photometer
became perfect for observing a single star with high precision. However,
the larger structures in the Galaxy, like open and globular clusters, could
not be investigated as a whole, but only from star to star. There was a
definite need for a new technology that mixed the precision of the
photoelectric photometer and the wide view of photographic plates. That was when the
{\bf CCD} (Charge-Coupled Device) appeared in astronomy, but only at the largest
astronomical centers. I myself started my scientific carrier in 1974 at Konkoly Observatory using
a Schmidt telescope and photographic plates taking pictures of an open cluster and later fields of
intermediate galactic latitudes to study their structure.
The photoelectric photometer had already been in use then, but the CCD
appeared in Konkoly Observatory two decades later than the 70's.

From this time the work on pulsating variable stars became fashionable. About nine times as 
many papers were published in this decade as in the previous one.
An obvious advantage of a star's pulsation started to be applied.
Pulsation was used to find out the physical parameters of the stars, and these were compared to
the observations. The investigations of {\bf Cepheids} were dedicated in the whole
decade to interpret the differences between masses deduced in different ways.
The determination of
absolute magnitude, color, effective temperature, and gravity allowed 
scientists to determine 
the masses ($\mathcal{M}_W$ - Wesselink mass) and evolutionary stages of 
variable stars. Using pulsation theory, three versions of the stellar
masses could be derived: from the pulsation constant ($\mathcal{M}_Q$), for bump Cepheids
($\mathcal{M}_{bump}$), and for beat Cepheids ($\mathcal{M}_{beat}$). The evolutionary theory
provided the evolutionary mass ($\mathcal{M}_{ev}$). 
The masses derived in different ways were supposed to be equal. \citet{Stobie1974}
reported ratios as $\mathcal{M}_Q$/$\mathcal{M}_{ev}$ = 0.70 and $\mathcal{M}_{bump}$/$\mathcal{M}_{ev}$ = 0.60 which
was not acceptable.
From time to time different ideas were suggested to resolve the discrepancy.
\citet{Fricke1971} argued that the plausible systematic
errors in the interior opacity, the distance calibration, or the $T_{eff}$ vs B-V
relation could cause a substantial error in the derived masses. A possible
mass loss was also mentioned by \citet{Stobie1974} as a cause of the discrepancy. It was generally emphasized
that an error in the pulsation model did not mean that the pulsation theory was 
incorrect, but rather that the error might be in the input physics. \citet{King1975}
suggested that self-consistent
masses might be obtained, if the color-temperature scale was used, which
reduced the Cepheid's effective temperature by 300-500 K below that normally
assumed. \citet{Iben1975} excluded the significant
mass loss in the extragalactic Cepheids. New opacities were calculated by
\citet{Carson1976}. Instead of the customarily used standard "hydrogenic"
Cox-Stewart opacities, the new opacities were based on the hot "Thomas-Fermi"
statistical model of all the elements heavier than hydrogen and helium.
The mass anomalies were suggested to be solved by the realization of a very 
helium-rich convective zone \citep{CoxAN1978}. The helium enrichment was supposed to be caused by
a Cepheid wind which blows away more hydrogen than helium, just as in the
solar wind. By the end of the decade the mass discrepancies were not 
completely resolved, but many of the discrepancies were been alleviated mostly
by an increase in the Cepheid luminosity and a decrease in their surface
temperature \citep{CoxAN1980}. The pulsation mass $\mathcal{M}_Q$ was in satisfactory agreement
with evolutionary masses ($\mathcal{M}_{ev}$) and the "Wesselink" masses ($\mathcal{M}_W$) were also fairly
satisfactory. Thus only the "bump mass" ($\mathcal{M}_{bump}$) and the "beat mass" ($\mathcal{M}_{beat}$) were anomalous with regard to evolutionary theory \citep{CoxJP1980}.
Probably inspired by the mass discrepancies, Cepheids were searched for 
binarity. A binary system allows for a precise direct measurement of the mass of the
members. 35 percent of a sample of 202 variables proved to be in a 
binary system \citep{Madore1980}. The evolutionary theory predicted
the change of the periods as the stars evolved. Classical Cepheids which
were monitored for a long time for light variability allowed researchers to check the
constancy or the change of the pulsation period.
\citet{Fernie1979} reported a linearly decreasing period at a rate of 8.06x$10^{-6}$
d/d for SV Vul, instead of the erratic period change that had been 
previously thought. Similar investigations were continued for many stars in the next decade. The mass discrepancy of $\delta$ Cepheids nicely shows that the observations are able to provide a guideline to improve the laboratory physics (opacity calculations). In addition, the changes of the parameters connected to pulsation (periods, amplitudes) have started to be investigated and which is a well-traceable method for the generation of space era.

The research on {\bf RR Lyrae stars} concentrated on three main lines, although they
were connected to each other: types of globular clusters, cause of the
period changes and the Blazhko effect. 
The original conclusion of \citet{Oosterhoff1939} claiming
that there were two types of clusters, was confirmed. In Oosterhoff I clusters
$<$ $P_{ab}$ $>$ $\approx$ $0.^d54$ (like M3, M5, M14), while for OoII type clusters 
($\omega$ Cen, M15)
the mean period of RRab stars were $<$ $P_{ab}$ $>$ $\approx0.^d64$ \citep{Stobie1971}. According to 
the author, the RRab stars in $\omega$ Cen were  more massive by 
$\Delta$log$\mathcal{M}$=0.10 or more helium rich by $\Delta$Y=0.20 than the variables in
M3. A difference in the mass loss was assumed. The metal content of the two
types were also different: OoI type was metal-rich Fe/H $>$ -1.0, while OoII type
was metal-poor with Fe/H $<$ -1.0 \citep{Butler1978}. \citet{vanalbada1973}
suggested that the Oostehoff dichotomy was caused by a dichotomy in the 
'transition period'. They argued that Christy's theoretical relationship
between the transition period and the luminosity could not be valid for clusters
of different Oosterhoff groups. The mode of pulsation of a star in the
transition region depended on whether it was evolving towards the left (OoI) or 
the right (OoII). \citet{Iben1970} interpreted that stars in M3 (OoI)
passed through the instability strip during the major phase of the core helium
burning and hence had high Z, whereas stars in $\omega$ Cen (OoII) passed the
instability strip rapidly from blue to red toward the end of the core helium 
burning and hence had low Z.

Cluster variables, the RR Lyrae stars were investigated for long enough up to this decade to check the
period changes of the variables in globular clusters of different type.
In early studies it was thought that the O-C diagrams were simply accumulated
evolutionary changes in the period \citep{Szeidl1975}. However,
the observations revealed a large variety of period changes. Different 
distribution of period changes were found for the 
globular clusters of two types, namely more stars showed period decreasing in 
OoI clusters and more period increasing were found in OoII type clusters 
\citep{Wehlau1975, Szeidl1973}. In addition, \citet{Iben1970} calculated the 
theoretical rate
of the evolutionary period change to be some orders of magnitude slower than
the rate of the period change observed for many of the cluster variables. They
inferred that most of an observed rate of the period change was 'noise',
confirming the suggestion of \citet{Balazs1965} that some stochastic
process was going on. Random mixing events occurring in the semiconvective
zone that could produce the observed period changes were mentioned by \citet{Sweigart1979}.
In a review paper \citet{Szeidl1975} came to the final 
conclusion that at that time it was impossible to attach any evolutionary
significance to the observed period changes. The investigation of time dependence became a general approach for each type of pulsating star with a long enough time base. The optimistic hope was to find the evolutionary time scale for the stars. Despite the discrepancies, it was useful to compare the theoretical evolutionary time scale to the observed one.

Furthermore, many RR Lyrae stars were known to show the Blazhko effect \citep{Szeidl1976}.
The observed period of this effect ranged from 12 to 537 days and a slow modulation
on a time scale of 4 to 10 years was found. There were RR Lyrae stars known where
the Blazhko effect entirely disappeared for many decades. The extreme
four year variation in the 41-day Blazhko effect of RR Lyrae itself was
supposed to be an intrinsic magnetic cycle in the stars.
 Beside the oblique
magnetic rotator explanation \citep{Balazs1962}, double mode pulsation was
suggested as a cause of the Blazhko effect \citep{Borkowski1980} through a mode 
coupling \citep{Dziembowski1982} of the fundamental mode to the second or third 
overtones. A decade passed and huge efforts were devoted finding a solution
for the mysterious Blazhko effect, but a consensus was not reached. If the funding policy of nowadays had been applied then, the investigation of Blazhko effect would have been stopped for a long time.


Considerable progress had been made in determining the multiple-period 
structure of individual {\bf $\delta$ Scuti stars} and identifying the excited radial
and non-radial modes through a variety of methods \citep{Breger1980a}.
Several groups were working on the problem all over the world, in the US in
Arizona (Fitch), in Texas (Breger), in Canada (Percy) and groups of Mexican, 
French, Italian and South African astronomers. By the end of the decade the observational efforts on
individual $\delta$ Scuti stars led to the conclusion  
that, like A-F stars in general, the $\delta$ Scuti stars were not completely
homogeneous \citep{Breger1980b}. The evolutionary status and the mass of the dwarf
cepheids were determined. The dwarf cepheids were supposed to be Pop I stars with
mass around 2 $M_{\odot}$ and a post main sequence evolutionary stage,
while SX Phe had low metallicity (Pop II), high space motion and low
luminosity. The $\delta$ Scuti instability strip could be 
schematically separated for sectors: the stars of high amplitude and
low radial velocity were found at the upper part, while the stars of low 
amplitude and faster rotation were at the lower part, on or near the main 
sequence. In classical Am (metallic line) stars pulsation was inhibited,
however, examples were found for pulsating metallic-line
giants, for the $\delta$ Del stars by \citet{Kurtz1979}. According to the Q pulsation
constant, the cool $\delta$ Scuti stars (on the right side of the instability strip in the HR diagram) tended to pulsate in fundamental mode, 
while the hotter ones (on the left side) pulsated in higher (1st, 2nd) overtones 
\citep{Breger1975}. 
The most outstanding example of the presence of the non-radial modes in the
light variation of $\delta$ Scuti stars was the case of 1 Mon, where modes
with equidistant spacings were found and identified as rotationally split 
modes \citep{Balona1980}. Period and/or amplitude variations were
reported for some stars by \citet{Fesen1973} and by \citet{Stobie1977} for 21 Mon.
However, the most well-known and extreme period/amplitude variation from night
to night was recognized by \citet{Stobie1976} in $\theta$ Tuc. Although
it was cited for years as a special type of $\delta$ Scuti star, it was obvious
that the co-existence of many non-radial modes could also cause an apparent
period/amplitude variation. To distinguish the two cases,  
properly distributed long data sets were needed. Finally, \citet{Kurtz1980} proved using an additional
observation of 70 hours over 21 nights that the amplitude of the main period
was stable over 7 years, but not the others. My personal connection to $\theta$
Tuc was that 20 years later, carrying out a three-site international campaign
we could prove that there were nearly 1 c/d frequency differences among three 
modes which could not be resolved from single site observations 
\citep{Paparo1996}.

Concerning the modeling of $\delta$ Scuti stars, it was \citet{CoxJP1976} who summarized the difficulties.
The non-radial pulsation seemed to be far more intricate than the corresponding
theory of purely radial oscillation, for this reason it was in a considerably
less-developed state. The fully non-linear, non-radial, non-adiabatic
calculation of stellar oscillations had not been attempted by anyone.
The full set of non-linear, partial differential equations that describe the
pulsation of stars had not been solved without significant approximations or
special assumptions for any star or stellar system. In the theory of non-radial
oscillations of a star it is customary to assume that each physical pulsation
consists of the product of a purely "radial" part, a function of the radial distance
alone, a spherical harmonic and an exponential time dependence. The order
of the harmonic is given by the value of angular harmonic index $\ell$, which is
positive integer or zero. Possible number of the other index m, are 2$\ell$+1. The
problem of mode identification (determination of the harmonic indices) for
non-radial oscillations in realistic stellar models had proved to be very
difficult.
With such assumptions, a detailed
linear survey of stellar models intending to represent dwarf cepheids and 
$\delta$ Scuti stars was carried out by \citet{Stellingwerf1979}. Growth rates, 
periods, and other parameters for the first six radial modes were given. The
nonadiabatic stellar modes indicated that non-radial modes were probably present
as well \citep{Stellingwerf1980}. However, the author emphasized that some basic
physical dissipation mechanism seemed to be missing. 
Due to the lack of the non-radial period ratios, the only possibility for identifying the non-radial modes was to compare the
observed period ratios to the theoretically derived radial period ratios. If
the observed period ratios did not agree with the theoretical radial period ratios, 
the modes were identified as non-radial modes.
This was reacting to the physics of the  model.
Hard efforts have been made to solve the problem of mode identification. 
\citet{Dziembowski1977} derived  formulas for the light and radial velocity 
variations of non-radially oscillating stars in the case of linear 
approximation. The author showed that the Wesselink technique, similarly to 
the case of radial pulsation, might also be used to determine the radius of
a star, if the spherical harmonic order of the oscillation is known, or
alternatively to determine the spherical harmonic order, if the radius is known.
\citet{Balona1979} extended Dziembowski's idea and formulated the
practical application of the Wesselink technique for this purpose.

Investigations in this decade resulted in the belief that pulsation and especially 
non-radial pulsation were ubiquitous among the stars. They occurred in $\beta$ Cep
stars, variable line profile B stars, $\delta$ Scuti stars, the white dwarf components
of some cataclysmic variables, the Sun, and the ZZ Ceti stars. Huge efforts
were invested to know more about their pulsational behavior.

\citet{Ledoux1951} showed that free radial oscillations could explain some of the
observed properties of the puzzling class of {\bf $\beta$ Cepheids}, however, he did
not consider what the cause of such oscillations were. Observations in UBV from the ground
\citep{Jer1971} and in the far ultraviolet from different satellites 
\citep{Hutchings1980} were carried out to know more on the pulsation
behavior, however, the cause of the oscillations remained unknown. 
\citet{Osaki1974} suggested a resonance between
the non-radial oscillation of the whole star and the overstable convection in the
rotating inner core. \citet{Aizenman1975} found a nuclear driven vibrational
instability in some of the lower $g^+$ modes. Nonlinear coupling was suggested 
as a mechanism by which this instability might excite the shorter-period f or
p, or perhaps radial harmonic modes which were supposed to be observed in the
$\beta$ Cephei stars. However, the final conclusion by the end of this decade
was that no viable instability mechanism had been found for their pulsation 
\citep{Lesh1978}. 

To complicate our understanding of B stars, \citet{Smith1977}
found pronounced changes in the line profile of {\bf 53 Persei}, which was located
far from the classical $\beta$ Cephei region. In the same year \citet{Percy1977}
reported small, but significant light variations for 53 Persei. This
star, the prototype of a new group, was the first star to show both photometric
and spectroscopic variations that could be best explained in the framework of
non-radial pulsations \citep{Buta1979}. Definitely much work will be done on
these stars in the next decades. I joined a multi-site photometric 
campaign on 53 Persei twenty years later to resolve the full 
frequency spectrum that could not be acquired from single site observations 
\citep{Huang1994}. 

After the discovery of the first
pulsating {\bf DA white dwarf} ({\bf ZZ Ceti stars}) by \citet{Landolt1968} during systematic surveys in
broad range of colors, it soon became clear that the ZZ Ceti variability was 
caused by pulsation. The instability strip was then defined and it was established that the
ZZ Ceti stars were otherwise normal white dwarfs. The variability was an evolutionary 
effect and all DA stars passed through the temperature range in which the
variables occurred \citep{McGraw1980}. The improvement of our knowledge on pulsating 
white dwarfs was fast because we can observe many pulsations of a ZZ Ceti
star per night. A great improvement in the realistic modeling of ZZ Ceti stars was 
also expected in the following decade. The investigation of the 5-minute oscillation
of the Sun became a distinct field, called helioseismology.

\section{Results between 1980-1990}

The new observational tool, the CCD, was slowly appearing in the smaller
observatories/institutes. New instrumentation was developed only on
specific request, like a multi-chanel photoelectric photometer for doing
fast photometry for compact objects, or a multi-slit photoelectric magnetometer
for measuring the circular polarization of 230 spectral lines simultaneously. 
The CORAVEL (CORelation-RAdial-VELocites) instrument started to operate at La Silla
in 1981.
Space instruments started to be used
mostly in the far ultraviolet (International Ultraviolet Explorer, IAU, 1978) and 
in the infrared (Infrared Astronomical Satellite, IRAS, 1983), which 
cannot be observed from the ground, due to the 
atmosphere of the Earth. The Hubble Space Telescope (HST) was launched in
1990, but due to technical problems it started the science mission only in 1993 after servicing.

In this decade not only the classical radially pulsating groups (RR Lyraes, Cepheids), but the non-radially pulsating groups ($\delta$ Scuti, white dwarfs)
were intensively investigated, not only observationally, but the new
observations immediately inspired extended theoretical work. Less improvement
was achieved for $\beta$ Cep stars, but the research status of a new group, 
the Long Period Variables (LPVs) is also mentioned in this decade.

The research of the {\bf RR Lyrae stars} in this decade started with two exciting 
papers based on new observations. The first was still connected to the Oosterhoff
types of globular cluster. \citet{Sandage1981} compared the period-amplitude, the period-rise time,
the period-color and the color-amplitude relations for M3 (OoI) and M15 (OoII) type 
clusters and  
recognized that all correlations of the parameters that involved the periods were
shifted by $\Delta$ log P = 0.055 toward longer periods in OoII type globular
cluster, whereas the other relations of the two cluster coincided. The only
consistent explanation was that the horizontal branch of M15 was brighter
than that in M3 by $\Delta$ log $L_{HB}$ = 0.090. As \citet{Smith1984} summarized
the series of Sandage's papers, the period of an individual cluster RR Lyrae 
star at a particular effective temperature decreased with increasing metal
abundance. RR Lyrae stars are intrinsically fainter in metal-rich clusters
than in metal-poor clusters. The RR Lyrae stars of different metallicity
might also have different luminosity. Of course, huge efforts were put
into determining the metallicity of individual RR Lyrae stars. \citet{Smith1984} used the
$\Delta$s method which was a means of determining the metal abundances from
low-resolution spectroscopy. It was hoped that if photometry of high accuracy
was available, the metal abundance of individual RR Lyrae stars could be 
determined from just their light curve. Definitely the space data of 
unprecedented accuracy offer the possibility to follow this suggestion and get
the determination of metallicity for numerous RR Lyrae stars. The statistical 
approach might reveal some, up to now hidden, structural connections.
Nevertheless, in the light of these results the question seems to be reasonable:
are RR Lyrae stars good standard candles?

The second interesting paper reported the discovery of ten double mode RR Lyrae
stars in the M15 (OoII) globular cluster \citep{CoxAN1983}.
The period ratio of the
fundamental and first overtone resulted in ($\Pi_1$/$\Pi_o$) = 0.746. Surprisingly,
the period ratio of RR Lyrae stars gave a reasonable mass, compared to
the double mode Cepheids, where the mass based on the period ratio was very
low. Many clusters were checked for double mode 
RR Lyrae stars and about three dozen were found by 1987. However, in the
$\omega$ Cen cluster no doubly periodic RR Lyrae star was found 
\citep{Nemec1986}, suggesting that the number of double mode RR Lyrae is
connected to the metallicity or the actual brightness of the horizontal branch.

Significant observational discoveries inspire immediate theoretical reactions,
as we saw in the previous decades. Of course, there was a large interest in
modeling the double mode
pulsation in RR Lyrae stars. It was shown that based on the simultaneous effect
of resonant mode coupling, the 2:1 resonance between one of the two linearly
unstable modes and a higher frequency mode caused double mode (fundamental and
first overtone) pulsation \citep{Dziembowski1984}. According to the 
authors, the resonant mode interaction might be an effective amplitude limiting 
mechanism in oscillating stars. \citet{Stothers1987}
concluded that second overtone pulsators probably did not exist among
RR Lyrae stars. Finally, stable multimode pulsations were shown
to be possible in state-of-the-art RR Lyrae models, but did not have periods close
to those observed in beat RR Lyrae stars \citep{Kovacs1988}.
The simultaneous effects of the resonant mode coupling and the nonlinear 
saturation of the driving mechanism were also studied by \citet{Moskalik1986}. Under 
certain 
conditions the only stable solution was a limit cycle in the form of a slow
periodic modulation of the amplitudes. In the case of RR Lyrae stars the 2:1
resonance between the fundamental and the 3rd overtone was the most likely cause
of such behavior. Upon a reasonable choice of parameters one could reproduce
observed periods of the Blazhko effect. A strong suggestion was that the
amplitude-modulated stars might be closely related to $RR_d$ stars and the same
mechanism might be responsible for both phenomena. 
A new cornerstone in the explanation of amplitude limiting mechanisms and the Blazhko effect is the idea of the resonant mode coupling.

A different observational/theoretical approach also appeared.
Fourier decomposition was
employed to compare the light curves of RR Lyrae stars with those emerging
from the hydrodynamical models. The $\phi_{21}$, $R_{21}$ and $\phi_{31}$ were
plotted against the period. The general definitions of the time-dependent
phase difference between the fundamental mode and the harmonics and the
amplitude ratio are given as: $\phi_{k1}$ = $\phi_k$-k$\phi_1$ and $R_{k1}$=$A_k$/$A_1$.The $RR_c$ type pulsators stand out from the 
$RR_{ab}$ stars, particularly on the $R_{21}$ plot which was found to be a
more sensitive discriminator of Bailey a, b, and c type stars \citep{Bailey1902} than the traditionally 
employed amplitude-period diagram \citep{Simon1982a}. Such kinds of investigations will be definitely even more effective on space data. A very good agreement
was obtained between theory and observation for the $RR_c$, but there were 
significant discrepancies in the Fourier phase quantities $\phi_{21}$, and
$\phi_{31}$ for $RR_{ab}$ \citep{Simon1985}. The calculated values were considerably larger than 
the values determined from observed stars. It was not clear how drastic a 
change would be necessary in the models to get an agreement. According to Simon,
a new series of hydrodynamic calculations was needed to answer the
remaining questions. In an effort to explain the discrepancy of period ratios
for Cepheids, \citet{Simon1982b} showed that arbitrarily increasing the opacity by a 
factor of 2-3 would solve the discrepancy. The new opacity code, OPAL, 
using an improved treatment of the atomic physics significantly increased the
Rosseland mean opacity of metals in astrophysical mixtures \citep{Iglesias1987}.
The new opacities confirmed what had been outlined as a need by observations over the previous decade. 
This is another data point to establish the time scale of the progress in pulsating stars.

Additional efforts were carried out in this decade for solving the mass discrepancy of
{\bf Cepheid variables} from both an observational and a theoretical point of view.
The new P-L-C relation was obtained by \citet{Schmidt1984}, which revealed a 
discrepancy in the absolute magnitude as much as 0.4-0.6 magnitude.  A new 
set of homogeneous Wesselink masses was obtained for 101 classical Cepheids
\citep{Gieren1989}. Theoretical models of normal Pop II type Cepheids in the
period range of 1-10 days were constructed with the Carson opacities 
\citep{Carson1981}. The derived masses closely agreed with masses determined
directly from atmospheric analyses and indirectly from stellar evolution theory. The theoretical luminosities predicted from the standard evolutionary tracks
had to be increased by 0.5 mag, but they were entirely consistent with the modern
evolutionary tracks that included some degree of convective core overshooting
\citep{Carson1988}. Observations of Cepheids in local and 
distant galaxies comprised one of the Key Projects approved for the Hubble Space
Telescope in the first cycle \citep{Simon1990}. Unfortunately HST was properly working
only from 1993. The main goal was to pin down the Hubble constant to an 
accuracy of $\pm$10 \% to determine the correct distance scale.
The nonlinear pulsational behavior of several 
sequences of Cepheid models was computed with a numerical hydrodynamic code
\citep{Moskalik1990}. These sequences exhibited a period doubling, as
the control parameter, the effective temperature, was changed. The period
doubling was caused by the resonance of the type (2n+1)$\omega_o$ $\approx$ 
2$\omega_k$ (n is an integer), and it turned out to be different
for Pop I (single doubling) and Pop II (a cascade of period doubling or
chaos) Cepheids. The period doubling is a phenomenon which becomes more 
important in the future.  

More and more additional details made our view of Cepheids colorful, however,
in most cases there was no consensus in their interpretation.
Much attention was paid to the so-called s-Cepheids as well which were classified
as low amplitude pulsators with short period and sinusoidal light curves. 
They were Pop I Cepheids that did not follow the Hertzsprung progression of the 
classical Cepheids, but had a progression of their own 
\citep{Antonello1990}. In the $\phi_{21}$- P plane they exhibited separate 
sequences. The short period group with P $<$ 3d had high $\phi_{21}$ values, while
the long period group with P $>$ 3d had low $\phi_{21}$ values. According to
\citet{Antonello1990}, both groups were pulsating in the first overtone and
they attributed the apparent sharp break in the $\phi_{21}$ diagram near 
3d to a resonance between the first overtone and a higher normal mode.
However, \citet{Gieren1990} argued that the low $\phi_{21}$ value represented
fundamental mode pulsation as was predicted by theory, only the light 
curves for some reason differed from those of the stars along the classical
fundamental mode sequence. The periods of the long period region essentially coincided
with the observed range of classical double-mode Cepheids \citep{Simon1990}.
As a result a far-reaching precedent started in the well-defined $\delta$ Cep stars and when new details appeared, apparently new subclasses were created. So, instead of unification we divided the HR diagram into smaller and smaller boxes.

The unusual Cepheid HR 7308, with the shortest known period was extensively 
investigated \citep{Breger1981}. The long-term follow up revealed that  the
amplitude varied by over a factor of 5 with a period of about 1210 days.
The amplitude variation suggested a Blazhko effect (observed in RR Lyrae 
variables, but not in Cepheids) or that an evolutionary event was taking place.
Radial velocity variation was monitored during three years with the spectrophotometer
CORAVEL, confirming the same periodicities as in the light curve 
\citep{Burki1982}. The type
of the excited mode was doubtful. According to \citet{Fernie1982} HR 7308 was not
an overtone pulsator. In contrast, an intercontinental simultaneous survey revealed that
HR 7308 was a classical Cepheid pulsating in the second (or higher) overtone. 
However, the possibility that the star was a Pop II Cepheid and/or that it was
pulsating in the fundamental mode or the first overtone could not be ruled out
definitively \citep{Burki1986}.

More and more precise rates of period change were obtained as the time baseline increased.
Secular period changes of 100 northern Cepheids were investigated with the help
of O-C diagrams \citep{Szabados1983}. The rates of the period changes were in good
agreement with those determined from stellar evolutionary theory. The frequency
of binary stars among the Cepheids was as high as 25 \%. Most of the
companions of Cepheids were B-type stars \citep{Szabados1982}.

Of course, the {\bf $\delta$ Scuti stars} being in the childhood of their
investigation and, due to their having shorter periods than the classical Cepheids, 
attracted more interest observationally.
Individual $\delta$ Scuti stars were intensively investigated. Different 
pulsational behaviors were reported: (i) a single sinusoidal component which 
did not agree with the fundamental mode \citep{Bossi1983}, (ii) a mixture
of radial and non-radial modes \citep{Bossi1982}, (iii) monoperiodic
pulsation in the second overtone \citep{Pena1981}, (iv) pure
radial pulsation in first and second overtones or fundamental and first
overtone with He depletion \citep{CoxAN1984} (v) mode switching
\citep{McNamara1984, Paparo1984}, (vi) multiple close 
frequencies \citep{Breger1987}, (vii) an amplitude variability on a time scale 
of years \citep{Breger1990a}, which was compared to the Blazhko effect of
RR Lyrae stars and (viii) a period decrease \citep{Breger1990b, 
Guzik1990}. The metallicity and the pulsation connection in $\delta$ Scuti 
stars got a new aspect.
Pulsation was also found in a {\bf classical Am star} \citep{Kurtz1989}.
Even the discovery of a new class of variable stars, the rapidly oscillating {\bf Ap (roAp)
stars} was announced \citep{Kurtz1982}. The amplitudes of oscillations were 
modulated with the same period and phase as the magnetic strength. This fact
suggested that the excitation mechanism of the oscillation in Ap stars was somehow
related to the strong magnetic field. All Ap stars were located in the $\delta$ Scuti
instability strip, therefore some physical connection between the pulsation of
$\delta$ Scuti stars and those of Ap stars was suggested \citep{Shibahashi1987}.
Despite the huge effort in investigation of $\delta$ Scuti stars, \citet{Kurtz1986}
called the attention to collaborative contemporaneous observations of 
multi-periodic $\delta$ Scuti stars from more than one observatory. He claimed this was
needed much more than the discovery of new $\delta$ Scuti stars or incomplete
frequency solutions for many stars. 

Concerning the theoretical work, a basic
step was done by \citet{Fitch1981}. He used a full stellar model and calculated 
linear adiabatic periods for all models. He presented 489 Q values for
35 modes. The survey was complete for the
range 0 $<$ $\ell$ $<$ 3. For double mode $\delta$ Scuti stars the period ratio resulted
in a serious discrepancy between the theoretical and the observed values. According
to \citet{Andreasen1988}, a helium depleted outer zone with a relatively
low mean molecular weight due to inward He diffusion might be the correct
explanation for the observed $\Pi_1$/$\Pi_o$ $\approx$ 0.77. However, there was a 
great discrepancy between the number of observed and theoretically 
excited frequencies.
Different concepts were checked for
getting an appropriate mechanism for the amplitude limitation and mode
selection. The effect of the parametric resonance on the development of 
acoustic mode instability in a model of Zero-Age Main-Sequence, ZAMS $\delta$ Scuti stars was 
first discussed \citep{Dziembowski1985}. Later amplitude equations
were derived describing the three-mode coupling in the presence of rotation
\citep{Dziembowski1988}. For the evolved $\delta$ Scuti stars with very
dense non-radial frequency spectra the phenomenon of mode trapping was
investigated as a possible mode selection mechanism 
\citep{Dziembowski1990}. There was evidence that this effect, rather
than saturation of the opacity mechanism, determined the mode amplitudes.
However, the nonlinear theory of stellar non-radial oscillation had not reached 
the level to enable the calculation of the amplitude of the unstable modes.
In any case, the prospect of asteroseismology rests on our
ability to connect measured frequencies to specific eigenmodes of stellar
oscillations. This would mean the possibility for mode identification which would be 
an important step in the asteroseismology of $\delta$ Scuti stars.
Mode trapping as a mode selection mechanism was first mentioned in this decade, but later we will find this explanation in connection with other type of pulsating stars.

New efforts were made in the determination of the spherical
harmonic degree $\ell$ and the azimuthal order m of a mode from observations. 
\citet{Balona1986a} proposed a simple method of analyzing spectral line profiles
in non-radially oscillating, rotating stars avoiding the profile modeling.
What was more, an algorithm was presented which allowed a completely objective
determination of all parameters for line profile variation in a non-radially
oscillating star \citep{Balona1986b}. \citet{Watson1988} presented histograms as a
function of $\ell$, among them was the amplitude ratio versus phase difference plane
($A_{col}$/$A_{\lambda1}$, $\Phi_{col}$-$\Phi_{\lambda1}$). Compared with
the model predictions, the light and color data seemed to assist in mode
discrimination for low $\ell$. \citet{Garrido1990} calculated the six possible
discrimination diagrams from the four Strömgren filters, and they found that
the diagrams v-y and b-y were discriminant between radial and low order 
non-radial pulsations, irrespective of the physical parameters such as 
temperature, gravity or pulsational constant for the $\delta$ Scuti models
considered. Although seemingly many ways of mode identification were available,
only a few modes in a few $\delta$ Scuti stars were identified. It was
predicted by theoreticians that data sets that were long and continuous were the
missing cornerstone of the massive mode identification in $\delta$ Scuti stars.
Observers in my generation often heard the following sentence: ``go to the 
telescope and observe more and the problem will be solved''. We will see that
it is not easy to identify modes in the non-asymptotic regime (p modes),
especially for fast rotating stars, even from the much longer, more
continuous space data that we only dreamed about in this decade.
Mode identification in stars pulsating in the non-asymptotic region is also a good indicator of the speed of progress. The problem was theoretically discussed in the last decade, but reliable practical applications are presented here. It will be followed also in the next decades.

The most impressive improvement was achieved in the field of {\bf white dwarfs} along
the lines of doing real asteroseismology. From the observational side, partly they were
ideal targets for single site ground-based observations 
\citep{Kawaler1989}, and the new astronomical instrument, the Whole Earth
Telescope (WET, \citet{Nather1990}) concentrated the efforts of existing
telescopes distributed in longitude to measure designated targets. The details
of the pulsational features grew incredibly quickly. The theoretical studies of ZZ Ceti (DAV) stars led to the prediction that the DBV white dwarfs 
should also pulsate driven by helium partial ionization zone. The same team
discovered the first DB pulsator, GD 358 \citep{Winget1982}. Soon, groups of
different ages were recognized: PNNV ($10^4$ yr), DOV ($10^5$ yr), DBV (
$10^7$ yr) and DAV (ZZ Ceti, $10^9$ yr) \citep{Winget1988}. According to the
rate of the period change (9.9x$10^{-15}$ s/s), a ZZ Ceti star (G117-B15A) proved to be 
one of the most stable clocks of any kind in the sky \citep{Kepler1988}, which
was equivalent to a 6.9 x $10^8$ year evolutionary cooling time scale. For the
prototype of the hot group, PG 1159-035, the rate resulted in dP/dt=-1.2x$10^{-11}$ s/s,
and a faster evolutionary time scale, $\tau$$\approx$ 1.4x$10^6$ yr \citep{Winget1985}. Impressively, evolutionary changes could have been detected in one to two 
decades using stellar pulsation.

Mostly multi-periodic oscillations were observed, but mono-periodic
pulsation (and its harmonics) was found in a DBV star 
(PG 1351+489, \citet{Winget1987}) that was both unanticipated and unexplained.
In addition, a significant pulsation frequency near 3/2 $\nu_o$ was 
localized, not only in this DBV star, but in DAV white dwarfs, one of which
was GD 154. Let me include a personal connection to GD 154. In the frame of an 
educational project for MA students 
I initiated the observation of GD 154 at Konkoly Observatory 
over a whole observational season (about 6 months) in 2006 (G. Fontaine 
suggested the star, due to the large amplitude). We confirmed
the surprising finding of \citet{Robinson1978}, that on the last
night the 3/2 $\nu_o$ became the dominant mode. In  2006 GD 154 showed
multimode pulsation, except at the end of the observing run, 
when the pulsation switched to monoperiodic pulsation \citep{Paparo2013}. 
Alternating high and low amplitude cycles and temporarily increased amplitudes
of some cycles (about 30 \% more pulsational energy) were 
presented in a poster even earlier, but the revulsion of the editor concerning the content led to the withdrawal
of the poster paper from the proceeding. Years later (suggested by P. Bradley), though, we did finally get 
these unusual behaviors published in the aforementioned paper.
We will see that space data revealed similar effects,  such as alternative 
cycles-period doubling for RR Lyrae stars and outbursts for white dwarfs. It is definitely a
great advantage for the present generation that they do not have to fight for
the acceptance of the observed data. Comparative analyses of
the observation of the subsequent years for GD 154 revealed that the star not only varied
its pulsational behavior between a "simple" multi-periodic and a quasi mono-periodic
phase with harmonic and subharmonic peaks, but there were differences between
the visibility of the near-subharmonics in the latter phase 
\citep{Bognar2019}. GD 154 will be a target of TESS (private communication by Zs. Bognár). GD 154 epitomises the situation that \citet{Winget1988} 
summarized in his white dwarf review: "we will find little that we do expect and
much that we would never expect". I think this statement will be extremely valid
in the space era. 

Nevertheless, there were other exciting results in this decade in which nobody doubted.
Two statistically mean period intervals, "characteristic period spacing" of 21.0$\pm$0.3 or 8.8$\pm$0.1 s were 
derived from observation for the hot white dwarf star PG 1159-035. 
A very good agreement with 
the theoretically derived period intervals resulted in the determination
of the mass (0.6 $M_\odot$) and the identification of the modes as $\ell$=1 or 3,
or both
\citep{Kawaler1988}. The determination of the physical parameters of white 
dwarfs meant
that the
modeling of the groups of white dwarfs was also a great success in this decade.
 The model had a simple stellar 
structure, C/O core, thin layers of helium and/or hydrogen depending
on the group. For the hotter groups (PNN and DOV) the partial ionization of
C and O were also taken into consideration. However, the problem of the selection
mechanism also occurred here. The resonance of a certain g mode to the thickness
of the surface composition layers meant that mode trapping could be 
a possible filtering mechanism \citep{Winget1988, Kawaler1990}
Convective blocking was reported as another candidate for a spatial filter
in ZZ Ceti stars \citep{Pesnell1987}. Color observations confirmed the theoretical
expectation that white dwarfs evolve at constant gravity \citep{Fontaine1985}.
The instability strip was narrow and all the white dwarfs showed pulsation as
they crossed the instability strip along their evolution. The research level of 
white dwarfs was so advanced that \citet{Winget1988} was able to raise the question that we wanted only 
to guess at the beginning of the space era: can we reach the asteroseismological level? He gave two possible 
answers, a pessimistic and an optimistic one. The pessimistic was that  
asteroseismology will never work, but the optimistic one was that it would give us 
the age of the Universe. Let us hope that the space data will take us to the
optimistic solution.
Despite the success of white
dwarfs \citet{Winget1988} suggested that non-linear non-radial calculations were
highly needed. Although these calculations are complex, supercomputers were
improving, but a basic step was real progress in understanding convection.
He called it the "Holy Grail" of hydrodynamics, since it was important in all 
helio- and asteroseismology.
White dwarfs are discussed in detail as an excellent example to show how important the ratio between the length of the pulsation period and the time base of the observations are. Due to the shorter pulsation period the progress is accelerated compared to the time dependence of progress in the classical pulsating stars or in the non-asymptotic pulsators. Mode trapping became an obvious explanation for mode selection in white dwarfs.

For many decades it was supposed that the {\bf $\beta$ Cep variables} were the only
pulsating B stars \citep{Balona1990}. The discovery of the {\bf 53 Per stars}, the moving bump
of $\zeta$ Oph and the periodic light and line profile variations of many
Be stars confused the picture. Although the theoretical problems in 
$\beta$ Cep stars were still existent \citep{CoxAN1987}, many observers moved to the 
observation of new features instead. Although 150 stars were classified as  
$\beta$ Cep candidates, even their definition as a group was not clear based 
on the type of pulsation \citep{Sterken1990}.
However, according to the observational H-R diagram, all $\beta$ Ceph stars were 
confined to a very narrow spectral range from B0.5 to B2. Their evolutionary 
stage could be close to the end of the core hydrogen-burning, the second 
contraction or the shell-hydrogen burning phase. The most prevalent view was
that practically all B stars were unstable to radial or non-radial pulsations
and that the $\beta$ Ceph stars represented merely an island of short (possibly
p-mode) pulsations in an ocean of long-period g-mode pulsation \citep{Balona1990}.

Advances in our knowledge of the details of long-period variability have been
gained rather slowly due to the extended nature of the atmosphere and the
dominance of the convective energy transport in the interiors \citep{Wood1987}.
The {\bf LPVs} occurred at a very interesting phase in the life of all the stars 
independently of mass, when they were most luminous and gave insight into their 
evolutionary mass loss.
The old LPVs belonged to a kinematic system which had a systematic velocity that
was indistinguishable from that of the H I gas \citep{Bessell1986}. Because
of their great brightness and the existence of the P-L relation, these stars might
be a useful group of extra-galactic indicators \citep{Wood1985}.

\section{Results between 1990-2006, before the space era}

Though data collection 
mostly continued in traditional ways, 
different approaches were followed to obtain more and more adequate data-sets
for solving specific problems. The success of the WET organization in the 
research of pulsating white dwarfs had a great influence on getting more
appropriate data for other pulsating stars, too. 
Some low-amplitude $\delta$ Scuti and roAp stars were also observed by the WET
organization. On the other hand, coordinated multi-site international campaigns were more and 
more often organized to get as continuous data as possible for $\delta$ Scuti
stars, $\beta$ Cep stars, especially for $\gamma$ Doradus and {\bf SPB} stars, where
the determination of the characteristic periods of the pulsation were highly disturbed by the
daily gaps. Most of these campaigns were organized by different individuals
from time to time. However, there were two permanent networks, the 
Delta Scuti Network (DSN) and STEPHI for solving the problems of $\delta$ Scuti
stars. Some hundreds of hours of data were collected and analyzed together. Another advantageous option appeared for the variable star research in this decade when
telescopes dedicated to special problems started to work. SDSS \citep{Ivezic2000} was a multi-color survey which aimed at providing a 3-D map of the Universe over a large part of the sky. The other projects,
MACHO \citep{Alcock1997}, OGLE \citep{Paczynski1994}, OGLE II \citep{Udalski1997},
EROS \citep{Beaulieu1995}, MOA \citep{Bond2001} photometrically monitored tens of millions of stars in different parts of the sky to search for dark matter
with micrelensing phenomena. These surveys happened to include variable stars, too. 
The new efforts led to new results, adding new mosaics to the general knowledge.
These surveys have extended archived databases, that could be also important in connection to the space data for checking time dependence on longer time scales.

The long-standing unsolved problem of the Blazhko modulation remained a central
issue, but on a higher level. The dedicated telescopes improved the database 
and new concepts appeared.
Amazing efforts were devoted to the interpretation of the Blazhko effect 
in {\bf RR Lyrae stars}. Both an extensive multi-site photometric observation
over a 421-day interval in 2003-2004 \citep{Kolenberg2006} and spectroscopic
coverage over an entire Blazhko cycle \citep{Chadid2006} of RR Lyrae
itself were carried out to make a choice
between the two possible explanations. According to the resonant model, the
non-linear resonant coupling of the dominant radial mode and the non-radial
modes predicted a triple frequency structure with side lobes of similar
amplitudes which were expected to be symmetrically placed. The magnetic model predicted
a quintuplet structure; however, in certain geometric configuration
only a triplet structure could have been expected. Both photometric and spectroscopic campaigns
resulted in a triplet structure in the frequency content, and there was no
sign of the quintuplet structure. It was hoped that high-quality and continuous space data 
will decide between the two possible explanations. Furthermore, the MACHO project \citep{Alcock1992}
incidentally gathered data for many Blazhko-type RR Lyrae stars.  It became clear
that the Blazhko effect had a different appearance.
There were Blazhko RRab stars with pure amplitude modulation or with both 
amplitude and phase modulation, or abrupt period changes could
occur. The Blazhko effect was found not only in RR$_{ab}$ stars, but in RR$_c$
stars \citep{Kurtz2000} as well. The existence of non-radial pulsation in a "pure"
radial pulsator had always been questionable. Finally,
non-radial pulsation  was reported in three RRc
stars in the M55 globular cluster \citep{Olech1999}. The theoretical 
calculations also clearly showed that low-degree non-radial modes could be 
excited  \citep{VanHoolst1998}. A large number of unstable modes 
in the vicinity of radial modes were partially trapped and presumably most
likely to be excited. However, their routine observations needed excellent 
photometric coverage, e.g. uninterrupted continuous space data.
The available high-quality ground-based data that had been accumulated for many RR Lyrae stars 
allowed the establishment of 
a relation between the metallicity parameter, Fe/H and the Fourier parameters, most importantly
$\phi_{31}$. It became possible to derive physical
parameters based on the morphology of the light curves for both RR$_{ab}$
\citep{Jurcsik1996} and RR$_c$ stars \citep{Simon1993}. 
A comparison of the
Blazhko-type and non-modulated RRab stars yielded the conclusion that there was
no Blazhko phase that seemed to be well-matched with that of non-modulated
RRab stars \citep{Jurcsik2002}. Definitely the RR Lyrae research was waiting
for the space missions.
The longer expected time base of the space data compared to the
length of the pulsation period or, more importantly, to the length of the period of the Blazhko effect, was expected to give remarkable improvement for the explanations.

The research for {\bf Cepheids} partly used the most advanced techniques and 
dedicated telescopes in getting more sophisticated data, at the same time
advanced theoretical investigations were carried out. 
Much progress was made in determining the fundamental properties of
Cepheids. The indirect determination of Cepheid masses seemed now to be in
agreement with the direct mass and the evolutionary masses due to the new opacities as
it was summarized by 
\citet{Percy1993}. The calibration of the zero point of the Cepheid P-L relation
was still a challenge, but CHARA and AMBER gave a good chance for the precise
determination of the stellar diameters and the related physical parameters
\citep{Moskalik2006}. The Hubble Space Telescope contributed 
to the binary Cepheids with a new discovery. Some previous binary systems were revealed to be
in fact members of a triple system. At least 44 percent of the Cepheids was
thought to be in a triple system \citep{Evans2005}. For many years AC And
\citep{Fitch1976} was the only triple-mode radial pulsator. Recently,
two Cepheids were discovered by OGLE in the Large Magellanic Cloud that
were pulsating in three radial modes \citep{Moskalik2005}.  {\bf From the theoretical side}, strange
modes were localized in radiative classical Cepheids and RR Lyrae models
\citep{Buchler2001}. These were surface modes predominantly trapped in 
the outer region of the star due to a very sharp and enormously high potential 
barrier in the partial ionization zone. The typical modal number of a strange
mode falls between the 7th and 12th overtone with a mmag amplitude level which
could be observed in the space era. These stars could be the "strange Cepheids"
or the "strange RR Lyraes". The discovery  of "anomalous" Cepheids (ACs), which had low metallicity but did 
not follow the P-L relation of Pop II Cepheids 
created some confusion in the identification of Cepheids. The new grid of 
evolutionary model by \citet{Fiorentino2006} led to the conclusion that the ACs
were the normal extension of Pop I classical Cepheids to lower metal content
and smaller masses. Hopefully, with more precise space data there will not be more and more separate subgroups, 
but that a unified explanation will be reached.
Comparing the different groups we can notice that mode trapping appeared in RR Lyrae and $\delta$ Cepheids, too, not only in white dwarfs. Once I was talking to a friend, who is a theoretician, about the similarities of the different type of pulsating stars, but at present independent codes are used for each type of pulsating stars. He immediately asked me whether
I was talking about a unified pulsation code, where the different types of
pulsating stars meant only different input parameters. I am an observer, so I may dream, although I know it is an extremely hard task. Definitely the first step would be a code for the non-linear treatment of nonradial pulsation.

The {\bf $\delta$ Scuti} research field was in the intensive data collection phase,
during which up to that time unexplained characteristics were recognized. The effect of
the chemical peculiarity on the pulsation got a new aspect, and the pre-main sequence {\bf $\delta$ Scuti stars} joined the company of the pulsating stars. Modeling of
time-dependent convection in $\delta$ Scuti stars had been started.
Much new data were collected for a great number of $\delta$ Scuti stars in this decade by 
coordinated multisite campaigns, although two permanent organizations
were the really successful ones. These were the Delta Scuti Network (DSN) founded by 
M. Breger, and STEllar PHotometry International (STEPHI) headed by E. Michel as
the coordinator. However, individual organizations were also  
successful (\citet{Paparo1996}, \citet{Paparo2000} for $\theta$ Tuc, and \citet{Paparo2018} 
for 38 Eri, based on a multisite campaign from 1998 and MOST observation from 
2011). The ground-based campaigns collected data in at least two colours, 
hoping for the possibility of mode identification. In the best cases,  
simultaneous spectroscopic observations were also carried out.
FG Vir, 4 CVn and XX Pyx became the most famous and best-studied $\delta$ Scuti 
stars in this decade. 
From 435 hours of observation 25 frequencies were derived for FG Vir in the last
campaign \citep{Breger1998}. Although the asymptotic relations did not apply 
exactly to $\delta$ Scuti stars, and the regularities were also invaded  by 
g modes and rotational splitting, regularities were searched for in FG Vir by the
histogram method \citep{Breger1999a}. XX Pyx was observed not only with 
DSN, but the WET, which happened to discover its pulsations. A different method, the Fourier Transform method was 
used for finding any regularity among the frequencies \citep{Handler2000}.
For 4 CVn an additional instrument, the Automatic Photoelectric Telescope (APT)
was also used over the years beside the DSN and WET runs \citep{Breger1999b}.
Amplitude variability was found on the time scale of years. 
In the majority of the well-studied $\delta$  
Scuti stars frequency pairs with 
less than 0.06 c/d (0.7 $\mu$Hz) frequency separation, as well as amplitude
variability were recognized \citep{Breger2002}. Mixed modes, trapped modes, mode coupling,
small spacing or rotational splitting were mentioned as explanations, but
the real reason was unknown. I participated in many DSN campaigns \citep{Breger1993, Breger1994, Breger1997, Breger1999}, but I was also involved in a 
STEPHI campaign on BS and BT Cancri
\citep{Hernandez1998}. To add more confusion in our view of the pulsation
in the $\delta$ Scuti instability strip, it was found that the chemically
peculiar $\lambda$ Bootis stars pulsated with high overtone modes 
\citep{Paunzen2002}. A multi-site campaign was organized for a pre-main 
sequence $\delta$ Scuti star \citep{Ripepi2006} to discriminate 
between the pre-main sequence and the post-main sequence evolutionary stage. The frequency
distribution in the low frequency range seemed to be a good criterion.
A promising step in the theory was that the red edge of the $\delta$ Scuti instability
strip for non-radial modes was obtained for the first time, using   
time-dependent convection (TDC) \citep{Dupret2004}.
The investigation of $\delta$ Scuti stars could have or at least is expected to have faster progress using space data.

A new class of pulsating stars, the {\bf $\gamma$ Doradus stars} were announced by
\citet{Balona1994a}. Very intensive observational efforts were dedicated to
find the connection between the new group and the chemically peculiar 
(Am, $\lambda$ Bootis) and $\delta$ Scuti stars \citep{Rodriguez2006,
King2006}. Especially the number of multi-site campaigns increased,
since the long period light variation made uninterrupted data necessary to determine unambiguous 
periods. I also joined some campaigns as a contributor \citep{Zerbi1997, Zerbi1999}. 
A variety of explanations appeared: some mentioned non-radial pulsation, others talked about star-spots or
quasi-stochastic amplitude modulation. Circumstellar dust shells were found
around Gamma Doradus itself \citep{Balona1994b}. The description of
$\gamma$ Doradus stars as a group was given by \citet{Kaye1999}. The stars 
typically had 1 to 5 periods ranging from 0.4 to 3 days with photometric
amplitudes up to 0.1 mag in Johnson V. The consensus of the light
variation was a high-order, low-degree,
non-radial gravity mode pulsation. After discovering 70 new $\gamma$ Dor 
candidates from the Hipparcos photometry it became clear that the location of the
$\gamma$ Dor stars on the HR diagram overlapped with the instability strip of
$\delta$ Scuti stars \citep{Handler1999}. The first convincing hybrid pulsator,
pulsating in both p- and g-modes, were reported by \citet{Henry2005}.
The reason of the excitation was reported soon after the identification of the 
group. The pulsations were found to be driven by the modulation of the 
radiative flux by convection of the base of a deep envelope convective zone
\citep{Guzik2000}.
The long period of pulsation biased by the alias structure of the ground-based observation pointed toward the need of space observation.

This period was a great decade concerning the improvement of our knowledge
of {\bf B stars} from both theoretical and observation respects.
The long-standing problem of the source of the excitation had been finally 
solved. 
The new opacities \citep{Rogers1992a, Rogers1992b} with the enhancement of heavy
element opacity by a factor of 2 to 5 over the Los Alamos opacity due to a 
large number of iron lines at the temperature of around T $\approx$ 2x$10^5$ degree 
highly contributed to finding the excitation mechanism in B stars 
\citep{Osaki1993}. Three groups were simultaneously working on the modeling of {\bf $\beta$
Cep} pulsation \citep{CoxAN1992,  Moskalik1992, Kiriakidis1992} and showed that the $\kappa$ mechanism acting in the
partially ionized zone of the elements of the iron group could account for the
pulsation of both radial fundamental and low-degree non-radial modes. 
\citet{Gautschy1993} confirmed the excitation of g-modes in stellar models
appropriate for early-type variables. Moreover, \citet{Pamyatnykh1999} found that
both short-period (low-order acoustic and gravity modes) and long-period
(high-order gravity modes) oscillations were excited in the same stellar model
of proper physical parameters.
Observationally the confusion on B stars of light and line profile variability became less serious. 
The $\beta$ Cep stars were accepted to pulsate in low-order p-modes in
accordance with many multi-site international campaigns (e.g. \citet{Handler2004, Handler2006}
 that I also contributed to). At the same time, a part of the 53 Per stars that surrounded the
instability zone of the $\beta$ Cep was found to belong to an independent
group of pulsating stars, named as {\bf Slowly Pulsating B Stars (SPB)} 
\citep{Waelkens1991}. The multi-periodicity and amplitude behavior unambiguously 
pointed toward pulsation in high radial-order of g-modes (n). (The low spherical harmonic degree (l) of these modes was given later \citep{decat2004}). Even 53 Per, the
prototype of the line profile variables was reported to be an SPB star 
\citep{Chapellier1998}. The multi-frequency pulsation of $\zeta$ Oph and its 
possible relation to $\beta$ Cephei variables was concluded from the long and
high precision photometry obtained by the Canadian satellite, MOST 
(Microvariability and Oscillations of Stars, \citep{Matthews2004, 
Walker2003}) and ground-based spectroscopy \citep{Walker2005}.
However, the distinct separation of the groups changed soon. One of the so-called
"classical" $\beta$ Cep stars turned out to be a $\beta$ Cep/SPB hybrid
pulsator, showing both low-order acoustic and high-order gravity modes that
was predicted by \citet{Pamyatnykh1999}. The word "hybrid" started to appear
more and more often in stellar pulsation. It predicts that the traditional
groups of pulsating stars are not as distinct as we thought. Maybe we are
moving to a unification of some, up to now separate groups.
The case of B stars shows that the progress of research highly depends on the advanced theoretical background.

{\bf Long-Period Variables} had been reported very briefly in this review, however, at the end of the
last century \citet{Barthes1999} highlighted that
the theoretical treatment of the structure, the pulsation and the evolution of 
Long-Period Variables (LPVs) was still imperfect. The efforts to improve it
suffered both from the complexity of the involved phenomena (strong convection
coupled to non-linear pulsation, thermal pulses, dredge-up of heavy elements,
mixing,  non-LTE, molecules, grains, shockwaves, high mass-loss etc), and
from the basic 
problem that the physical parameters were not known. However, the 
gravitational microlensing surveys (MACHO, OGLE, OGLE II, EROS, MOA) 
revolutionized this field partly because 
their strategy of observing a large collection of stars at large distances let
us estimate the luminosity of these stars \citep{Fraser2006}. It became
well-established that pulsating red giants were located on a series of up to six
parallel Period-Luminosity (P-L) sequences \citep{Wood2000}. LPVs first pulsated
in overtone mode (1st to 3rd) and switched to fundamental mode pulsation when
having crossed some luminosity limit. For the first time the measurement of the masses of
AGB stars had shown that mass loss of the order of 0.3 M$_\odot$ occurred on
the RGB and AGB \citep{Lebzelter2006}.

It would be hard to separate the observational and theoretical results for the
{\bf white dwarfs}, since the unprecedented progress in the development of white
dwarf interior structure models allowed researchers to fit them immediately to the 
available observations \citep{Metcalfe2005}. This rapid improvement in the 
understanding has been most evident for the DBV white dwarfs, where the
physical conditions were ripe for asteroseismic investigations. It became clear
that the detailed core C/O profile and a double-step helium were the most important features for  
quantitative asteroseismology of the DBV white dwarfs \citep{Dehner1995}. The seismic study of the 
interiors of white dwarfs fed back into the theory of stellar evolution 
\citep{Kawaler1998}. Hot white dwarfs were also mature examples of active 
asteroseismology with such a great importance that in their interiors their prior evolutionary history, 
especially the story of their last nuclear evolutionary stage had been
locked. 
Seismological data had demonstrated the action
of gravitational settling in white dwarf envelopes and provided measurements
and important upper limits to the cooling rates of white dwarf stars 
\citep{Corsico2004}.
The observed cooling rate of DAV stars proved to be in good agreement
with the theoretical predictions \citep{Kepler2005}, however, in other types of
white dwarfs, (DOV and DBV) the observed period changes were faster than the
theoretically predicted values whose reason was unknown. 

Exciting results for white dwarfs included determinations of stellar
masses, distance, rotation rates (both surface and interior) and internal
stratification of composition were reported. \citet{Kawaler1994} derived that 
the constant period spacing was primarily determined
by the mass of the star, which allowed mass determination. Regular departures from 
uniformity could have been regarded as a result of mode trapping by a surface composition discontinuity
\citep{Kawaler1998} allowing the determination of the mass of the H and He
layers. Rotation could be deduced by identifying the equal frequency split in
the pulsation spectrum.  Different amount of rotational split from multiplet to
multiplet suggested the departure of the 
interior from solid-body rotation. 
The observed splits in GD 358 (DBV type) indicated that the outer layers of this
star rotated much faster (by a factor of nearly 2) than the inner regions \citep{Winget1994}.
\citet{Kawaler1999} explored the possibility 
to estimate the whole internal rotation
profile. This has been one of the final goals of asteroseismology. 
The 
possibility of the crystallization in the core of white dwarfs had been
discussed. The discovery of a DAV lying near the red edge of the instability strip
and having characteristic long periods, but anomalously low amplitudes 
provided the first opportunity to search for the 
observational signature of crystallization in an individual star \citep{Kanaan1992, Kanaan2005}.
\citet{Montgomery1999} suggested that the core  of this star might be up
to 90 percent crystallized, depending on its mass and internal composition.
The crystallization process led to one of the largest sources of uncertainty
in the ages of cool white dwarfs \citep{Segretain1994}.
In an optical spectroscopy approach two ZZ Ceti (DAV) stars were discovered 
\citep{Fontaine2003a}, which brought the number of known ZZ Ceti stars to a 
total of 34. They concluded that the instability strip was pure, in which no 
non-variable star was found. A pure instability strip was mentioned at the 
first time for a group of pulsating stars.
The number of discovered non-radially pulsating white dwarfs in cataclysmic
variables was rapidly increasing by the aid of SDSS \citep{Nilsson2006}.
It is worthwile to emphasize that the phenomenon of mode trapping became a useful method in determining the physical parameters of white dwarfs, in not more than one or one and a half decades.

The second new group of pulsating stars in this decade, the {\bf pulsating subdwarf B (sdB) stars} were discovered in a surprising way, allowing a new field for
doing asteroseismology.
In the frame of the Edinburgh-Cape (EC) Blue Object Survey
a small-amplitude, very rapid light variation was discovered in the sdB
binary, EC 14026-2647 \citep{Kilkenny1997}. It was regarded as a prototype
of a completely new class of variables which was referred to as EC 14026 stars
(also called sdBV$_r$ or V361~Hya stars).
The main period was 144 s with a semi-amplitude of $\approx$ 0.012 mag with
a possible second period near 134 s. The variations were more obvious
at U than at V which was a strong indicator that variation originated in the
hot sdB star rather than in the cooler component, a main sequence F or early
G type companion. Kilkenny's discovery inspired a search for other stars
of this kind. EC 14026 sdBV stars were found with six periodicities 
\citep{Koen1997}, with periods showing evidence of a variable amplitude 
\citep{Stobie1997} and with evidence for period/phase changes in the main
period \citep{ODonoghue1997}. Interestingly, at the same epoch, the 
existence of pulsating sdB stars was independently predicted from theoretical
considerations based on the identification of an efficient driving mechanism
for the pulsation \citep{Charpinet1996}. The driving mechanism was due to
an opacity bump associated with heavy-element ionization. It was found that the
models showed low radial order unstable modes; and both radial and non-radial
(p, f and g) pulsations were excited. They felt confident enough to risk the
prediction that some subdwarf B stars showed luminosity variations resulting
from pulsational instability. This was one of the best examples when inductive
(observation) and deductive (theoretical) investigations met! However, the
sequence of the discoveries was not finished with the short-period hot sdB
stars. During the course of an ongoing CCD monitoring program a new class of
low-amplitude, multi-mode sdB pulsation with periods of the order of an hour 
was discovered \citep{Green2003} and named originally {\bf lpsdB} (or {\bf sdBV$_s$ or V1093~Her stars} in the new nomenclature \citep{Charpinet2011}) as a group. 
The periods were 
more than a factor of 10 longer than those of EC 14026 stars implying that they 
were due to gravity modes rather than pressure modes and they were located
on the cooler part of the Extended Horizontal Branch (EHB). The same $\kappa$ 
mechanism successfully explained the presence of longer periods, however, 
radiative levitation was needed to boost the iron abundance in the driving 
region and modes with $\ell$=3 and/or 4 were excited, not the "more visible" $\ell$=1 
and/or 2 modes \citep{Fontaine2003b}. Almost naturally, examples of the hybrid 
hot and cool sdB stars, pulsating both in short and long periods, were 
discovered \citep{Oreiro2005, Schuh2006}. In addition, pulsation
was discovered in a {\bf He-rich subdwarf B (He-sdB) star}. According to the periods,
the pulsation was more likely due to a high order, non-radial g-mode than to
radial or non-radial p-modes \citep{Ahmad2005}. Using thirteen 
simultaneously excited periods in a hot sdB led to the first 
asteroseismological determination of the fundamental parameters, such as the star's 
temperature, its surface gravity (with a greatly improved accuracy), its total
mass, and the mass of its H-rich envelope \citep{Brassard2001}. The first
hot pulsating sdO star was discovered by \citet{Woudt2006}. The relatively rapid
discovery  of many new pulsating groups required a systematic order in the
nomenclature \citep{Kilkenny2010}.  A summary of the names is given in a table, but
the basic concept of the changes is that rapid, slow and hybrid pulsations are
marked by "r, s and rs". Asteroseismology can be applied in the future 
in a new part of the HR diagram. Could new changes in the opacity lead
to discovery of new groups of pulsating stars? There seems to be not much
space for new groups on the HR diagram. Started from  the classical instability strip,
almost all part of the HR diagram is covered now by pulsating stars, except for the coldest part of the main sequence, the M and brown dwarfs, and the most luminous part of the HR diagram.
 
\section{Newest results of the space era, 2006-2018}

Space-borne instruments are the new generation of the technical
developments. The Canadian MOST (Microvariability and Oscillation of Stars, 
launched in 2003, ended in 2014) space telescope was the first spacecraft 
dedicated entirely to the study of asteroseismology 
\citep{Matthews2003, Walker2003}. In the last few 
years the research field of variable stars has benefited more and more from the 
operation of space telescopes, although their primary goal was the detection of 
exoplanets. The French-led CoRoT (Convection, Rotation and planetary Transit,
launched in 2006, ended in 2013) and the NASA-built Kepler telescope (launched
in 2009, ended 2012, but operated as the K2 mission between 2013-2018) were joint
missions for planetary and stellar research \citep{Baglin2006, Borucki2010}. 
Searching for exoplanets, a huge amount of data was gathered together
for the stars that were located in the fields of view.
The newest American space telescope, TESS (Transiting Exoplanet Survey
Satellite, launched in 2018) produces the first-ever spaceborne all-sky transit
survey to detect small planets with bright host stars in the solar neighborhood
\citep{Ricker2015}. The upcoming ESA telescope, PLATO (PLAnetary Transits
and Oscillations of stars, with a planned launch in 2024) will devoted to figuring
out under what conditions planets form and whether those conditions are
favorable for life \citep{Rauer2016} or not. The Swiss-led ESA mission, CHEOPS (Characterizing
ExOPlanet Satellite), which is ready to be launched in 2019, is planned for
the study of the formation of extrasolar planets \citep{Rando2018}. With so many dedicated 
instruments, it is not surprising that planetary science has become as 
fashionable nowadays as variable star research used to be some decades ago. However,
one has to admit that variable  star research also benefits from the purely 
exoplanet-oriented projects.
Beside the exoplanet missions, a basic space telescope, the GAIA, was launched in 
2013 which was designed to measure the position and distances of 
stars with unprecedented precision 
\citep{Eyer2013}. The Gaia data release 2 was recently published \citep{Brown2018}.
Definitely the greatest advantage of space telescopes for the presently 
active generation of scientists is that they do not have to suffer for collecting data on
individual objects from night to night. In addition, space-based 
data are continuous with different lengths (6 months for CoRoT and 4 years
for the Kepler mission) and the precision is now at the unprecedented ppm level.
There is good hope to answer some, many, or all the questions that have been asked
in the last decades. Due to the fact that the data are more easily available,
and there is publication pressure for getting financial support, 
the number of published papers has increased enormously. International teams are
organized for working on the data of a given mission and the same teams migrate from one space mission
to the next one. The structure of the national institutes seems to have been loosened
and scientific connections with colleagues outside of those permanent teams
have been reduced/or vanished. The whole research field has been revolutionized, which is not
a disadvantage in itself. However, here are some thoughtful points. The results of certain space missions are presented in specialized conferences (CoRoT weeks, KASC, TASC and BRITE meetings). I am sure that in the future we will have TESS, GAIA, CHEOPS and PLATO meetings, too. On the other hand, thematically highly specialized meetings are organized more and more often (RR Lyrae, white dwarf and sdB meetings, although white dwarf meetings are not only for pulsating white dwarfs, but for the whole community who are working on white dwarfs in any respect.) Smaller workshops are also organized for mostly the international teams. Such a great change in the policy of presenting the scientific results compared to the fewer IAU supported general meetings in every three years may have
some unplanned consequences. Although we still have the IAU supporting more general meetings, unfortunately the financial support is not enough to attend both the general and the specialized meetings. The knowledge of the experts is getting deeper, but concerns a more and more narrow field, especially if we consider the flood of published papers. There is no time to properly follow the whole field of pulsating stars, only a smaller and smaller section of the general knowledge. So, another important question for the space era is whether we are able to synthetize/assimilate such a large flood
of data. Do we have enough time and energy to theoretically interpret
the new discoveries, beside the very convenient task of data treatment? The
future will tell.

It is impossible to cover all results obtained from space data up to now, rather 
I intend to give a 
taste of what we could answer from the previous questions, what we still have to
solve or what the new fine details are that appeared in connection with pulsating 
stars.

The past shows that temporary confusion seems to appear when new data are
obtained, especially if we do not immediately have guiding principles. The
advanced ground-based instruments, the space data of long time base and
of unprecedented precision questioned the constancy of the observable
parameters of the pulsation in {\bf Cepheids}. Just to name a few, amplitude irregularities, period fluctuations,
doubled periods, non-radial pulsations and multimode radial pulsations were
found.
Although the Irish Astronomical Journal based on the paper
of \citet{Grand1954} reported that the "structure of the atmosphere of the 
Cepheids is on solid ground", a program on the secret life
of Cepheids (SLiC) is nowadays going on in which the aspects and behavior of
classical Cepheids, which are still regarded as unclear, were planned to be studied
\citep{Engle2014}. Observations were obtained both from the ground and space. 
Due to the application of space instruments (HST Cosmic Origins Spectrograph in
UV and XMM-Newton, the X-ray Multi-Mirror Mission) and the improvement in 
precision, even the prototype of the classical Cepheids, $\delta$ Cephei, 
surprised us. For example, high temperature plasma ($10^3$ - $10^7$ K) was detected above 
the Cepheid photosphere with variable X-ray activity which was attributed to 
pulsation-driven shocks propagating through the Cepheids' outer atmosphere and
giving rise to mass-loss through a stellar wind. The HI 21 cm line 
observations with the Very Large Array (VLA) directly determined a mass of
circumstellar atomic hydrogen $\mathcal{M}_{HI}$ $\approx$ 0.07 $M_\odot$ with a 
mass-loss rate of d$\mathcal{M}$/dt $\approx$ (1.0$\pm$0.8) x $10^{-6}$ $M_\odot$ yr$^{-1}$
\citep{Matthews2012}. Modeling showed that the combination of moderate 
convective core overshooting and pulsation-driven mass loss could solve the Cepheid
mass discrepancy \citep{Neilson2011}. In addition, the $\delta$ Cephei
proved to be a spectroscopic binary based on measurements of high-precision radial
velocities \citep{Anderson2015}. Using all available data, the orbital period 
was 2201 day, the eccentricity, e=0.647, and the companion mass was constrained 
within 0.2 $M_\odot$ $\le$ $\mathcal{M}_2$ $\le$ 1.2 $M\odot$. The close pericenter 
approach of the two stars has far-reaching consequences for the explanation
of the observed circumstellar environment of $\delta$ Cephei.
The MOST
photometric measurements revealed low-amplitude irregularities in Cepheid stars.
The nearly continuous high-precision photometry also revealed alternations in
the amplitudes, the first case of a period doubling detected in a classical 
Cepheid. The period doubling led to the appearance of half integer frequencies
\citep{Molnar2017a}. The Kepler data disclosed significant cycle-to-cycle 
fluctuations in the pulsation period, indicating that classical Cepheids might
not be as accurate astrophysical clocks as commonly believed 
\citep{Derekas2012}. However, this statement does not seem to be valid overall. 
Using MOST data, a fundamental mode pulsator had a
light curve that repeated precisely, an overtone pulsator on the other hand
showed light curve variation from cycle to cycle \citep{Evans2015}. Concerning
the type of pulsation, CoRoT light
curves of Galactic Cepheids did not show any convincing
evidence of excitation of non-radial modes \citep{Poretti2015}. However, in the
Large Magellanic Cloud additional variability was discovered by the OLGE team
for Cepheids with period ratios in the P/P$_0$ = (0.60, 0.65) range.
A power excess at half the frequency of the additional variability was also
reported \citep{Smolec2016}. In addition, two triple mode classical Cepheids 
pulsating simultaneously in the first three radial overtones were also 
discovered \citep{Moskalik2009}. Hopefully, the details will sooner or later be
combined into an overall explanation.

The staunchest community of astronomers are those who have been working
from decade to decade to solve the mystery of the Blazhko effect.
Huge efforts have been devoted to getting ground-based data. 
The Konkoly Blazhko Survey
\citep{Jurcsik2009}, the Antarctica project \citep{Chadid2014}, the OGLE project
\citep{Smolec2015}, and the Catalina Sky Survey \citep{Torrealba2015} represented
such successful ground-based efforts, while the CoRoT, the Kepler and the GAIA space 
missions have provided a larger and larger contribution to the improved knowledge on
{\bf RR Lyrae stars}. The number of stars showing RR Lyrae characteristics has
increased incredibly, and definitely surprising statistical results have been
predicted in connection with the evolution of stars and the structure of our
galaxy. However, at present our picture has become much more confusing compared
to the simple categorization of RR$_{ab}$, RR$_c$, RR$_d$ subtypes that
we used to have in earlier decades. The discovery of additional low-amplitude modes
blurred the line between the three main groups \citep{Molnar2016}. Since
the first discovery of the additional frequencies that did not belong to 
the classical multiplet structure of a Blazhko-type RR Lyrae and was 
interpreted as a non-radial mode with a ratio of $f_0$/$f_{add}$ = 0.6965
\citep{Chadid2010}, all unusual frequencies were called additional 
frequencies. Nevertheless, these additional frequencies could have different
types, as well as definitely different physical origins,
although most of them were explained by some kind of resonance of the
higher-order radial modes. The period
doubling phenomenon was first detected in Kepler RR$_{ab}$ stars 
\citep{Kolenberg2010, Szabo2010} and later in CoRoT RR Lyrae stars,
\citep{Szabo2014}. The period doubling manifested itself as alternating
maxima and minima of the pulsational cycles in the light curve, as well as
through the appearance of half-integer frequencies located halfway between
the main pulsation period and its harmonics in the frequency spectra. According
to the theoretical explanation, the period doubling was caused by 9:2 resonance
between the fundamental mode and the 9th-order radial overtone showing
strange-mode characteristics due to mode trapping \citep{Kollath2011}. This was
one type of additional mode. The phrase ``additional modes'' was used 
for frequencies which could have been identified as the second radial overtone,
but the excitation of the second
overtone was not theoretically expected \citep{Poretti2010, Benko2010}. 
In this case, the group of $RR_d$ stars could be extended with stars
 where not the
fundamental and first overtone, but the fundamental and the second overtone were
simultaneously excited. Without any knowledge on the
amplitude of non-radial modes, these additional modes could be simply regarded
as a non-radial mode excited near the second radial overtone, where excitation 
was theoretically predicted. This was the second 
type of ``additional modes''.  
An additional frequency was reported as early as 2007 \citep{Gruberbauer2007} in
an $RR_d$ star.
The $f_0$/$f_i$ = 0.9272 ratio suggested that the additional mode was  
a member of the second type.
The third type of additional modes were the f$_X$ or
0.61-type (for the frequency ratio) modes which were strongly connected to the first overtone and could
be detected in both first overtone (RR$_c$) and double-mode (RR$_d$) RR Lyrae
stars \citep{Moskalik2015, Netzel2015}. Such modes occurred in Cepheids, too,
with a slightly different frequency ratio (0.60-0.64), as it was mentioned 
earlier. 
According to the theoretical explanation, the additional periodicities
arose from the harmonics of non-radial f-modes effectively trapped
in the outer part of the envelope \citep{Dziembowski2016}. In Cepheids the
modes had angular degrees from $\ell$=7 to 9, in RR Lyrae only $\ell$=8 and 9.
It seems that all cases represent the fact that non-radial modes were
discovered in the traditional radial pulsators, the Cepheid and RR Lyrae
stars. Unfortunately, the approaches do not agree on the definition of the
old/new subtypes. \citet{Molnar2017b} kept the traditional categories,
but colored them by the non-radial modes, while \citet{Netzel2015} kept the
traditional radial pulsation categories and created a new category for the
radial-non-radial-mode pulsators.
Hopefully, the picture will become clearer
in the future, but the final solution can only be given by a non-linear
nonradial pulsation code which can provide the amplitude ratio of the radial and
non-radial modes. The non-linear, non-radial code was already urged as early as
1988 by D. Winget in connection with white dwarfs \citep{Winget1988}. The 
successful progress of the asteroseismology of the different types of non-radially
pulsating stars (and as we see the classical radial pulsators also belong to 
them) requires further steps in modeling.
The newest results for $\delta$ Cep and RR Lyrae stars show that the space data will discover many new details on the classical radial pulsators, partly suggesting that they also belong to the non-radially pulsating stars, and only the amplitude ratio of the radial and non-radial modes are different compared to the rest of the non-pulsating stars. It strongly points toward the need to know more about the amplitude of the excited modes. The half integer frequrencies, the period doubling, and the trapped modes echoes the case of white dwarfs.

There has been no consensus on the reason of the {\bf Blazhko-type modulation} of the
RR Lyrae stars either, but the number of possible causes has been much reduced. 
A possible suggestion was that the Blazhko effect is caused by the combined effect of the fundamental and first overtone 
shocks on the atmosphere \citep{Gillet2013}. A beat
mechanism in double mode RR Lyrae stars were also suggested \citep{Bryant2015}. 
The more
widely accepted explanation was the resonant model \citep{Buchler2011}.
The suggested reason was the 9:2 resonance of the fundamental mode and the
high-order (9th) radial overtone, as in the case of the period doubling. The amplitude
equation formalism showed not only the period doubling, but also that the amplitudes 
were modulated, and that in a broad range of parameters the modulations were 
irregular. Observations showed that the Blazhko cycles did not repeat regularly 
\citep{Kolenberg2011, Guggenberger2011} and the Blazhko modulation
was multiperiodic \citep{Sodor2011, Benko2014}. In a remarkable
case, two different modulation periods were identified with a similar strength and
a different period ratio from season (5:4) to season (4:3) \citep{Sodor2011}.
One of the most important recent observational findings was the limited modulation
{\bf in the K band} \citep{Jurcsik2018}. They showed that the Blazhko modulation was
primarily driven by changes in the temperature variation and the radius variation
played only a marginal role. This empirical fact alone drastically reduced the
possible mechanism that could be responsible for the modulation. That was the
first observational evidence that the Blazhko phenomenon was strongly related
to the top of the atmospheric layers of the stars that was predicted by Christy in 1966 \citep{Christy1966}. \citet{Kollath2018} checked
the aforementioned models and \citet{Stothers2006} model versus the constraints.
According to the conclusion, there was no existing model which satisfied
all the observational constraints of the Blazhko modulation exactly. However, the
9:2 resonance hypotheses of the fundamental to 9th overtone at least did not
contradict any of the observational facts. An overview of explanations of the Blazhko effect for the past 60 years has been presented. The space data and the huge ground-based efforts led to more details, but there is no theoretical explanation that
is accepted by everybody. This is the longest that a problem has not been solved in stellar pulsation.

Space data increased our knowledge on the pulsation of  
{\bf $\delta$ Scuti stars} remarkably and let us answer many long-standing questions, but the validity of the 
new information changed from year to year, as more and more data were processed.
The $\delta$ Scuti instability strip seemed to contain both pulsating and
non-pulsating stars in about equal fractions. The amplitude distribution of the
frequencies did not support the suggestion that all stars pulsated
with amplitude of low level \citep{Balona2011a}. However, a different
sample and other aspects led to a conclusion that the instability strip was pure, 
unless pulsation was shut down by diffusion or another mechanism, such 
as interaction with a binary component \citep{Murphy2015}. In a different vein,
the precise
GAIA luminosities allowed the instability strip of $\delta$ Scuti and $\gamma$ Dor 
stars to be redetermined, which partly proved that less than half of the 
stars pulsate and that the two instability strips overlap \citep{Balona2018}. However,
the question why there are non-pulsating stars with roughly similar parameters
in the instability strip remains open.
New theoretical calculations confirmed the overlap of the two instability strips
and suggested that there was no essential difference between the $\delta$
Scuti and $\gamma$ Dor stars, they were only two, a p-mode and g-mode,
subgroups of one boarder type of pulsating stars below the Cepheid instability
strip, and  most of the variables were hybrids \citep{Xiong2016}. Consequently, 
it was not surprising that low
frequencies were present in at least 98 \% of $\delta$ Scuti stars 
\citep{Balona2018} and it was meaningless to continue to call them as 
$\delta$ Sct/$\gamma$ Dor hybrid stars \citep{Balona2015a}. However, it had been
reported that pure $\delta$
Scuti stars with no significant frequencies in the $\gamma$ Dor region (below
5 $d^{-1}$) did, albeit rarely, exist \citep{Bowman2018}. Although the largest part of two
instability strips overlapped both observationally and theoretically, 
the red and blue edges did not perfectly agree \citep{Balona2018}. In the $\delta$ Scuti instability
strip several peculiar types of pulsators were located. According to the precise 
luminosities, \citet{Balona2018} concluded that {\bf High Amplitude Delta Scuti (HADS) stars} were normal $\delta$ Scuti stars and
not transition objects and SX Phe stars were also located well inside the 
instability strip. Surprisingly, the roAp stars also exhibited $\delta$ Scuti and
$\gamma$ Dor pulsation \citep{Balona2011b}. Part of the low frequencies 
in $\delta$ Scuti stars
were attributed to different co-rotating surface features. Most of the Am stars 
had light characteristics of rotational modulation due to spots 
\citep{Balona2015b}. Flares were also detected in about two percent of Kepler
A stars \citep{Balona2013}. The surface metal enrichment in the peculiar stars
was explained as the result of diffusion and gravitational 
settling in the absence of a magnetic field. Two spectropolarimetric 
measurements revealed a clear magnetic structure in the Stokes profile in
a Kepler $\delta$ Sct/$\gamma$ Dor hybrid candidate \citep{Neiner2015}.
This was the first magnetic main sequence $\delta$ Scuti star. A weak magnetic
field was also detected in 1 Mon, the famous multi-periodic high-amplitude
$\delta$ Scuti star \citep{Baklanova2017}. The interaction of  
diffusion-pulsation and the physics of the stellar envelopes should 
be re-examined to see how a magnetic field could be generated in A stars \citep{Balona2015b}. Investigating
Kepler $\delta$ Scuti stars, 603 stars exhibited at least one pulsation mode
that varied significantly in amplitude over 4 years \citep{Bowman2016}. In
more general, amplitude and/or frequency variations had been found among
nearly all types of non-stochastically excited pulsators \citep{Guzik2015}.
The long-term follow-up of the famous $\delta$ Scuti star, 4 CVn, showed that
most pulsation modes exhibited a systematic significant period and amplitude changes
on a timescale of decades \citep{Breger2017}. The reasons of the
period and amplitude variability will definitely be investigated as space data on 
a longer 
and longer time base will be available in the future. Two kinds of anomalous 
pulsators seemed to exist between the blue edge of the $\delta$ Scuti and the 
cool edge of the $\beta$ Ceph instability strip (Maia variables) and between
the cool edge of the SPB and the blue edge of the $\gamma$ Dor instability
strips (hot $\gamma$ Dor variables) \citep{Balona2016}. The current models
did not predict pulsation in this region of the HR diagram. Nevertheless, a large 
population (36 stars) was found in this region in a young open cluster 
\citep{Mowlavi2013} and 17 similar stars were also identified in another
young open cluster \citep{Lata2014}. Both the rapid rotation and a revision
of the opacities were mentioned as a possible solution for these anomalous 
pulsators. It seems that explanations of the many new discoveries need serious
theoretical work on the different parts of the input physics.
Despite the space data of unprecedented precision the question of why some stars with the same atmospheric and physical parameters pulsate and others do not, has been still an unexplained issue for 60 years. Presumably this problem is also
connected to the amplitude limiting mechanism, which also remains unsolved.

{\bf Mode identification in $\delta$ Scuti stars} was still an unsolved problem,
due to the lack of simple, easily identifiable frequency patterns; especially
for fast rotating stars, it was a formidable challenge \citep{Reese2018}.
Nevertheless, we might say that 
remarkable improvement had been achieved in the last decade. Great
effort has been put into the establishment of the pulsational model of 
fast rotating stars \citep{Roxburgh2006, Lignieres2008, Lignieres2009}. 
\citet{Reese2008} suggested a new asymptotic formula for the island modes
which were later confirmed and for which theoretical echelle diagrams were built 
\citep{Ouazzani2015}. Realistic multi-colour mode visibilities were 
calculated for possible mode identification \citep{Reese2017} and were 
applied for BRITE data \citep{Reese2018}. The mode identification needed
data at least in 3 photometric bands and also required a large number of
acoustic  modes, preferably in the asymptotic regime. 
Unfortunately, it is hard to satisfy these requirements for the bulk of $\delta$
Scuti stars.
From an observational side, a quasi-periodic pattern was found in $\delta$ Scuti 
stars at first for CoRoT data, and a spacing periodicity around 52 $\mu$Hz was
derived \citep{Garcia2009}. Calibrated for $\delta$ Scuti stars
in eclipsing binaries, a scaling relation ($\Delta\nu$-$\rho_{mean}$) was 
established between the frequency spacing obtained from the pattern in the star
and its mean density \citep{Garcia2015}.
This relation was theoretically derived by \citet{Suarez2014}. Later the 
scaling relation was used to get the precise surface gravity without any constraints
from spectroscopy or binary analysis \citep{Garcia2017}. The scaling relation
could lead to the possibility of a massive statistical investigation of $\delta$ Scuti stars.
Not only the determination of the spacing, but the identification of independent
modes were aimed at searching for sequences of quasi-regularly distributed
modes, allowing for a tolerance level in the exact spacing 
\citep{Paparo2016a}. Echelle diagrams were constructed for 90 CoRoT 
$\delta$ Scuti stars and the characteristic spacings were derived 
\citep{Paparo2016b}.
Unfortunately, a test investigation revealed that the echelle diagrams could be 
built not only for the consecutive radial orders of a certain $\ell$ degree mode, 
but for the combination of eigenmodes and rotationally split modes, producing 
sometimes more than one spacing value. We could not disentangle
the normal eigenmodes from rotational splitting. However, a new observable, the shift 
between the sequences was derived. The fact that some
shifts between the sequences were integer multiples of the rotational split might 
suggest a possible resonance effect between eigenmodes and rotation as a 
selection mechanism \citep{Paparo2018}. 
Using 1860 CoRoT $\delta$ Scuti stars, a common regular pattern was found in 
agreement with island modes featured by theoretical non-perturbative treatment 
of fast rotation \citep{Michel2017}.
Regularities in the amplitude spectra of these
stars produced ridge-like structures with a spacing of order of a few tens of
$\mu$Hz (of order a few $d^{-1}$) which was consistent with the consecutive radial order 
pulsation modes. The f$_{min}$ and f$_{max}$, as well as the large separation
values  might be used as seismic indices to characterize stars, which opens
the perspective for ensemble seismology using $\delta$ Scuti stars.
The space data introduced progress in the identification of modes in the non-asymptotic regime, but there is room for the next generation to work on this field.

Although the {\bf $\gamma$ Doradus stars} were newly identified, the interpretation
of their pulsation seemed to be an easy task.
Spectacular improvement was achieved on the asteroseismology of g-mode 
pulsators on the main sequence, both on $\gamma$ Dor and SPB stars, as well as on the hybrid stars.
The asymptotic theory predicted regular period spacing patterns which was
evinced both in CoRoT and Kepler stars. A series of g-modes of the same 
$\ell$-degree and different radial orders n consisting of 24 frequencies were
found in a $\gamma$ Dor/$\delta$ Scuti hybrid \citep{Chapellier2012}. In addition,
strong coupling of the g-modes to the radial fundamental mode was also manifested. 
Rotationally split g-mode triplets and surface
p-mode triplets were discovered in a Kepler star \citep{Kurtz2014}. This gave the first robust 
determination of the rotation of the deep core and the surface of a main-sequence
A star. They showed with high confidence that the surface rotated slightly
faster than the core. Period spacings were determined
and echelle diagrams were constructed for several $\gamma$ Dor stars \citep{Bedding2015}. Small
deviations from regular period spacing were found that arose from the gradient
in the chemical composition just outside the convective core. 
\citet{VanReeth2018} presented an extended theoretical overview on the restoring forces for
rotating stars. The simple, exactly regular period spacing predicted by
the asymptotic theory is valid only for a non-rotating chemically homogeneous
star. The real manifestation of the period spacing depends on the rate of
the stellar rotation and in consequence, on the type of the restoring force 
(buoyancy, Coriolis or both). 
For slowly pulsating stars (buoyancy) the g-modes split into 
triplets, however, for moderate to fast rotators (the rotation rate is 20 \%
or more of the critical rotation rate) purely internal pulsation such as
r-modes (Coriolis force), or gravito-internal pulsation modes appear (both type
of forces are acting).
For stars pulsating in gravito-internal modes the period spacing
has a positive or negative slope depending on the sign of the azimuthal number, m (retrograde or prograde). 
In the frame of ensemble modeling of $\gamma$ Dor stars near-core 
rotation rates were determined from the observed period spacing pattern \citep{VanReeth2016}. 
For most stars gravity or gravito-internal modes were identified, but for the first time
Rossby modes were also found. A new observable,
the slope of the period spacing when plotted as a function of the period was 
derived which was uniquely related to the internal rotation 
\citep{Ouazzani2017}. The global Rossby waves (r-modes) were found to be 
present in many $\gamma$ Dor stars, spotted stars and heartbeat stars (highly
eccentric binary stars), and even in a frequently outbursting Be star. The common
feature of r-modes in the amplitude spectra was the presence of broad humps
that appeared immediately below the rotational frequency \citep{Saio2018}.

The {\bf SPB variables} were also identified as a separate group only in the
last decade, they shared the remarkable success of the  $\gamma$ Dor stars
along the way to asteroseismology due to the g-mode pulsations. The detection
of numerous gravity modes in a young star was reported from CoRoT data 
\citep{Degroote2010}. 
The mean period spacing allowed researchers to estimate the
extent of the convective core, and the clear periodic deviation from the mean
constrained the location of the chemical transition zone to be about 10 \%
of the radius. The first detailed asteroseismic analyses
of a cool SPB star, showing a series of 19 quasi-equally spaced dipole modes,
was reported from four years of Kepler photometry \citep{Papics2014}. The 
amount of splitting showed an increasing
trend towards longer periods which pointed toward a non-rigid internal
rotation profile. Independent modeling was done by \citet{Moravveji2015} 
and \citet{Triana2015}. In a next example, the longest unambiguous series of 36
gravity modes of the same degree $\ell$ with consecutive radial order n, which
carried clear signatures of chemical mixing and rotation was discovered
\citep{Papics2015}. According to the authors, this star should be considered as the
Rosetta stone of SPBs for future modeling. Over the years of the Kepler era the number of 
in-depth analyzed SPB stars in the original Kepler field doubled from four to
eight, and the number of SPB with an observed period spacing from four to nine; seven of these are from Kepler \citep{Papics2017}.
Simultaneously, it had been shown that the period series were not only common,
but they dominated the frequency spectrum of SPB stars.
Both the $\gamma$ Dor and SPB variables show that the results of high level observations can be straightforwardly interpreted if the theoretical background is available (the asymptotic theory).

The impact of MOST, CoRoT and Kepler and K2 missions on seismology of {\bf $\beta$
Cephei stars} was rather modest, because few of these stars were observed and
because one color photometry alone allowed mode identification only
through the recognition of eventual patterns in the stellar pulsation 
\citep{Handler2017a}. Several bright stars of $\beta$ Cep-type were observed by the 
WIRE satellite tracker, several by MOST and only one by CoRoT \citep{Degroote2009}. The largest
impact on the number of known $\beta$ Cephei stars came from the All Sky
Automated Survey (ASAS) \citep{Pigulski2009}. The BRITE Constellation 
(BRIght Target Explorer) was designed for asteroseismic studies of 
$\beta$ Cep stars. Despite the observational disadvantage, some results were
obtained both on $\beta$ Cep and $\beta$ Cep/SPB hybrids by combining space
and ground-based data. One hybrid pulsator had a sufficiently large number
of high-order g-modes and low-order pressure (p) and mixed modes were 
detected to be usable for in-depth modeling \citep{Handler2009}. In a triple
system with two massive fast rotating early B-type components, both components 
proved to be $\beta$ Cep/SPB hybrids. In addition, the system's secondary star had
a measurable magnetic field \citep{Pigulski2016}. In a 
$\beta$ Cep star 40 periodic signals were detected intrinsic to the star
with some of the previously known pressure and mixed modes and some newly
found gravity modes. Temporal changes in the amplitudes
were also detected. The disagreement of the observed and the theoretically  
predicted amplitude behavior might lead to incorrect identification, if using data
in optical filters only \citep{Handler2017b}. Kepler data showed that there
were non-pulsating stars in the $\beta$ Cep and SPB instability strip 
\citep{Balona2011}. Magnetic fields seemed to not be common in SPB, $\beta$
Cep and Be stars, although there were some magnetic B pulsators 
\citep{Silvester2009}. A huge effort was dedicated to getting precise physical 
parameters of pulsating stars with LAMOST (Large Sky Area Multi-Object Fiber 
Spectroscopy Telescope) \citep{DeCat2015}. Despite the massive efforts, the
models calculated with standard opacity tables could not explain the observed
oscillation spectra \citep{Briquet2011, Handler2012}. An opacity 
increase of a factor 2 at the depth at which nickel had a significant 
contribution would solve the discrepancies between the observed low-frequencies
and the theoretical high-order g modes \citep{Dasz2017}. Does it mean that we
still need to increase the opacity for getting a reliable model for $\beta$ Cep
variables?
The results obtained on space data for both $\beta$ Cep and $\delta$ Scuti stars point toward the urgent need for theoretical improvement of non-asymptotic pulsations. It is an extremely hard problem for the upcoming generation. However, they have to step further, otherwise the space data will be only described, but remain unexplained from a theoretical point of view.

The {\bf emission B (Be) stars} were not well-studied in the last decade. 
However, all space missions also observed Be stars, partly in dedicated runs, 
such as the MOST mission, but mostly as a by-product of their observing strategies 
(such as the CoRoT, Kepler, and also the solar mission SMEI - Solar Mass Ejection Imager). 
To date MOST has observed five Be stars, for several weeks each, Kepler has observed three 
stars for four years, CoRoT has observed 40 Be stars for between a few weeks and half a 
year, and SMEI almost 130 stars over nine years \citep{Rivinius2017}. From a
stellar pulsation perspective, Be stars are rapidly rotating SPB stars, that
is, they pulsate in low-order g-modes \citep{Rivinius2016} (precisely in low degree (l) and high order (n) g-modes \citep{decat2004}).  The BRITE
and SMEI joint data revealed that next to rapid rotation, non-radial 
pulsation seemed to be the most common property of a known Be star. 
Four large amplitude frequencies were exhibited. Two of them were closely
spaced frequencies of spectroscopically confirmed g-modes near 1.5 d$^{-1}$,
one slightly lower (about 10 \%) exophotospheric (Stefl) frequency, and at
0.05 d$^{-1}$ the difference frequency between the two non-radial g-modes 
\citep{Baade2018}. The circumstellar (Stefl) frequency did not seem to be
affected by the frequency difference, which underwent large amplitude variations.
Its variability seemed to be the main reason for the modulation of the 
star-to-disk mass transfer. When a circumstellar disk was present in a Be star, 
the power spectra were complicated by both cyclic, or
periodic and aperiodic circumstellar phenomena, possibly even dominating the
power spectrum \citep{Rivinius2016}. Depending on the spectral type, four 
categories of variability were distinguished, namely, 1) stochastic 
\citep{Neiner2012}, 2) bursting and 3) cleanly pulsating \citep{Semaan2013}, and 
4) almost harmonics \citep{Diago2009}. Optical light curves for 160 Be stars
obtained by the KELT (Kilodegree Extremely Little Telescope) and simultaneous
infrared and visible spectroscopy were analyzed to study the disk creation 
process and to monitor the evolution and demise of these disks once formed
\citep{Lab2018}. The duration of disk build up and dissipation 
phases were measured for 70 outbursts showing that the average outburst took
about twice as long to dissipate as it did to build up in optical photometry.

Progress in the field of {\bf white dwarfs and sdB stars} was even more impressive 
in the space era than the last decades. New versions of the stellar/binary
evolutionary endpoints were recognized as a new type of white dwarfs. The 
{\bf extremely low-mass (ELM) white dwarfs} having five pulsating members for the 
time of space era  \citep{Hermes2013}, represented a new group of white 
dwarfs with their 0.18-0.2 M$_\odot$ mass as opposed to normal white dwarfs with a mass of  
$\approx$ 0.6 M$_\odot$ and massive white dwarfs with a mass of $\ge$ 0.8 M$_\odot$.
Binary evolution was the most likely origin in the process of which they 
harbored very thick H envelopes and were able to sustain residual H nuclear burning 
via a pp-chain, leading to a markedly longer evolutionary timescale 
\citep{Corsico2016}. An ELM white dwarf was shown to be in a compact, 5.9-h 
orbit binary with a fainter, more massive WD companion, and the system exhibited 
both primary and secondary eclipses in the light curve \citep{Hermes2014}.
The discovery of other new group, several white dwarfs with atmospheres 
primarily composed of carbon with little or no trace of hydrogen or helium, was
reported \citep{Dufour2007}. The first pulsating member in the {\bf hot DQ group},
as the new group was named, was revealed by \citet{Montgomery2008}. 

Extended ground-based surveys were performed to find white dwarfs and sdB
stars in the Kepler field, and finally the surveys were successful. 42
white dwarfs were discovered in the original Kepler field \citep{Greiss2016},
and 27 white dwarfs were measured through the K2 mission \citep{Hermes2017a}.
The most surprising results in the DA white dwarfs were the detection of a 
new phenomenon, the large-amplitude outbursts at timescales of much longer than the 
pulsation period \citep{Hermes2015}, which was first reported on the
1.5-year long light curves of the first ZZ Ceti star discovered by Kepler 
\citep{Bell2015}. As the number of similar cases increased, it was 
established that only the coolest pulsating white dwarfs within a small
temperature range near the cool, red edge of the DA instability strip exhibited
the outbursts \citep{Bell2016}. There was discrepancy between the 
theoretically and observationally determined locations of the red edge of the
DA instability strip \citep{VanGrootel2012}. The outbursts
recurred stochastically on days-to-weeks timescales and could brighten a white dwarf
by more than 40 \% for several hours \citep{Hermes2017b}. In addition,
the fastest rotational rate (1.13$\pm$0.02 hr) of any isolated pulsating white 
dwarf known to date was found in K2 data. The highest mass (0.87$\pm$0.03 M$_\odot$) that was measured
for any pulsating WD with known rotation suggested a possible link
between high mass and fast rotation. The mean flux increase corresponded to
nearly 750 $^o$K \citep{Hermes2017c}. As soon as a DB white dwarf was found
in the surveys, it was immediately submitted for follow-up space observation.
Five modes roughly equally spaced in period were found with a mean spacing of
37 s. The three strongest modes showed a triplet with a mean splitting of
3.3 $\mu$Hz \citep{Oestensen2011}. In an other DBV, clear signatures of
non-linear effects were found that could be attributed to a resonant mode
coupling mechanism, which might motivate further theoretical work to
develop the non-linear stellar pulsation theory \citep{Zong2016}. The
hottest helium-atmosphere white dwarf known to pulsate exhibited a rich 
oscillation spectrum of low-order g-modes with clear patterns of rotational
splitting from consecutive sequences of dipole and quadrupole modes. These 
modes were able to probe the rotation rate with depth in this highly evolved 
stellar remnant. A bright spot was also recognized and used for the measurement
of the surface rotation rate (10.17 hr) \citep{Hermes2017d}. The space era
has been a very productive data collection phase for white dwarfs. Hopefully,
immediate theoretical work will be inspired to incorporate the newly
observed phenomenon into the white dwarf models.

The search for members of the new instability strip on the Extreme Horizontal Branch (EHB) continued and five {\bf sdO pulsators} were discovered in $\omega$ Cen \citep{Randall2011, Randall2016}. A very rapidly pulsating sdO star was discovered in the
Edinburgh-Cape (EC) survey with strongly variable amplitude, which may be a 
field analogue of the $\omega$ Cen sdO variables \citep{Kilkenny2017}.
The harvest for the {\bf sdB stars} based on space data started with the clear identification of nine
compact pulsators and a number of interesting binary stars 
\citep{Oestensen2010}. Using only 30.5 d of nearly continuous time series
from Kepler, more than ten independent pulsation modes were found in a 
short-period pulsator, and one longer periodicity showing that this sdB might
be the hottest member of the hybrid sdB stars. Additional periodic changes 
suggested that a significant number of additional pulsation frequencies might be
present \citep{Kawaler2010}. A statistical investigation was
done for 13 sdB stars concerning the nearly equally spaced periods. It was concluded 
that period spacings could be easily or readily detected and they are useful for
mode identification. Most of the stars indicated modes with $\ell$=1 and some showed modes with
both $\ell$=1 and 2 degree \citep{Reed2011}. It was amazing that a 2.75-year Kepler observation
with a 93.8 \% duty cycle containing 1.4 million measurements and resulting in
0.017 $\mu$Hz resolution was obtained for the most slowly rotating sdB star 
(88$\pm$8 day). Back then it was not surprising that 278 periodicities were found
in which 59 \% had been associated with low-degree ($\ell$$\le$2) pulsation modes.
According to the authors, this star represents a 'solved' sdB pulsator 
\citep{Reed2014}. We could not have issued such a statement before the space era!
More fascinating results were obtained: (i) the clear indication of mode trapping
in a stratified envelope \citep{Oestensen2014a}, and (ii) modes without long-term
coherency showing the stochastic characteristic of these modes beside the two
normal pulsation modes \citep{Oestensen2014b}. Such solar-like pulsations,
although suspected
in sdB stars, have never been observed before. Extremely complex
systems were also found. The near-continuous 2.88 year Kepler light curves
revealed  that one sdB star had an unseen white-dwarf companion with an orbital
period of 14.2 days. A rich g-mode frequency spectrum with a few p modes at
short periods was also found. The g-mode multiplet splittings constrained
the internal rotation period at the base of the envelope to 46-48 days as
a first seismic result for this star. The few p-mode splittings might point to
a slightly longer period further out in the envelope of the star, suggesting
the possibility of radial differential rotation \citep{Telting2014}. The
presence of two nearly Earth-sized bodies orbiting the post-red-giant hot
sdB star was also reported \citep{Charpinet2011}. The distances from the star were 0.0060 and 0.0076 AU and
the orbital periods were 5.7625 and 8.2293 hours, respectively. They were 
interpreted as the dense cores of evaporated giant planets that were 
transported closer to the star during the engulfment. 
Maybe we are closer to the optimistic prediction of Don Winget \citep{Winget1988} 
who said that
with asteroseismology we might be able to provide the age of the Universe.
Understandably, due to the physical limits of this publication, certain excellent results in
the successful field of pulsating stars had to be omitted.

Despite the great improvement in the asteroseismology of white dwarfs, 
$\gamma$ Dor and SPB stars showing pulsation in the asymptotic regime excited
by $\kappa$ mechanism on any kind of ionization zone, and even 
helioseismology, these cases represent only certain evolutionary stages. To
have asteroseismology for all evolutionary stages, we need improvement for the
non-asymptotic regime and the non-linear treatment of the non-radial pulsation 
in any kind of pulsating variables, in Cepheids, 
RR Lyraes, $\delta$ Scuti and $\beta$ Cep stars. We would need at least to
reach the possibility for ensemble asteroseismology, to get more precise
physical parameters and characteristic similarities for as many stars as 
possible to step further in seismic modeling. Space observations of
{\bf red giants (RGB stars)} present a worthy example.
 
In the stars on the main sequence that are similar to our Sun and also in the more 
evolved red giants, which represent the future of our Sun, small-amplitude 
oscillations are intrinsically damped and stochastically excited by the 
near-surface convection were predicted, which were sensitive to the
physical processes governing their interiors \citep{Brown1994, 
Chris2004}. \citet{Brown1991} suggested that the frequency
of the maximum oscillation power ($\nu_{max}$) might be expected to be a fixed
fraction of the acoustic cut-off frequencies, which is a directly accessible
seismic parameter. The spacing between the consecutive radial orders, e.g. the
$\Delta\nu$, was related to the acoustic radius and therefore to the mean
density. A scaling relation was derived by \citet{Kjel1995a}. The 
first case of a star other than the Sun showing an unambiguous evidence of
solar-like oscillations was published \citep{Kjel1995b,
Chris1995}. Several ground-based observations were
carried out in getting radial velocity measurements  
\citep{Frandsen2002, DeRidder2006} and photometry \citep{Stello2007},
which were followed by early space detections (WIRE - \citep{Buzasi2000},
HST - \citep{Stello2009}, MOST - \citep{Barban2007}, and SMEI - \citep{Tarrant2007}).
The real harvest started when the   
unprecedented precision of the space data (CoRoT, Kepler) allowed for the detection of the 
low-amplitude light variations {\bf in red giants}. \citet{DeRidder2009} 
presented the first study of 300 red giants observed by CoRoT showing that
they exhibited non-radial oscillations with common patterns. \citet{Hekker2009}
demonstrated that there was a tight relation between the large separation $\Delta\nu$ and
$\nu_{max}$ for {\bf red giants as in the} case of solar-like stars. This
opened the possibility to study the characteristics of their oscillations in a
statistical sense, unlike the traditional goal of asteroseismology where
the accurate measurement of individual mode frequencies was compared to the
model frequencies \citep{Huber2010}. \citet{Miglio2009} identified the signature of the red clump and
were able to distinguish the stars of the red clump from the red giant branch (RGB) stars, which
represent a different evolutionary stage. In the RGB stars helium burning
has not started in the core, while in the red clump stars started burning helium 
in the core. This is a defining moment in the life of a star
from an evolutionary point of view.
\citet{Kallinger2010} exploited the possibility of measuring stellar masses and
radii from asteroseismic measurements, even when the effective
temperature was not accurately known. In the following years larger and larger
samples were analyzed together, up to thousands of stars. The second cycle of the important
discoveries was the measurement of core rotation rates in subgiants and
red giants \citep{Beck2012, Deh2012}. These rotation rates
were slower than the current models predicted \citep{Marques2013}. Another 
mechanism is needed that can transport the angular momentum between the core and
the envelope \citep{Belkacem2015a, Belkacem2015b}. Less success was found in fitting stellar
models directly to the observed frequencies due to poor modeling of
the near surface layers \citep{Ball2018}. Fine tuning was carried out on the
scaling relations, namely, the empirical scaling relation was
based on more refined theoretical assumptions \citep{Kjel2011}. What is  
more, the non-linearity of the scaling relation was also discussed 
\citep{Kallinger2018}. The astrometric distances provided by
GAIA proved to be in excellent agreement with the asteroseismic distances 
\citep{DeRidder2016}. The overwhelming success of the seismic indices
inspired the preparation of the
Stellar Seismic Indices (SSI) database that intended to provide the scientific 
community with a homogeneous set of parameters characterizing solar-type
pulsators observed by CoRoT and Kepler \citep{deAssis2018}.
It is impossible to cite all the papers in a review with
limited space which evidently has to lead to excluding certain scientific work. 
This review only intended  
to indicate that the research field has improved a lot owing to the appearance of 
space 
data and inspiring immediate theoretical interpretation, which is
the ideal process of the scientific research.

\section{Conclusion}

In a review paper on the current status of asteroseismology \citet{Kjel2001}
predicted that "it is impossible to imagine that asteroseismology will ever 
reach a level similar to that in which we find helioseismology today 
concerning analysis techniques, data quality and the level of results". Maybe,
thanks to space missions, the excellent data of high precision and the
results obtained up to now, the future of asteroseismology is brighter than the
vision was before the space missions.
The data are easily available for everybody who is interested in the research 
field. However, for some space missions a priority phase is applied for those who financially supported the mission, before the data becomes public. It means a limitation for the countries which are financially not powerful. Of course, the easiest reachable results, discoveries and the predictions with prepared algorithms can be obtained first in the priority interval. To go deeper in the data requires a longer time and much more effort. As this overview shows, even in the past there was a delay of about 10-20 years from the technical point of view between scientists from different countries. Hopefully, enough members of an enthusiastic young 
generation will be ready to look for new aspects, and for new relations to give
answers to the long-standing questions and to draw up new questions in
the field of stellar pulsation beside the fashionable exoplanet research. The computing power and observational data are much more readily available worldwide now.

The present and forthcoming space missions will definitely give a strong
base for statistical investigations concerning the spatial distribution of stars
of a certain evolutionary status. The astrometric measurements will lead to a
better understanding of how our galaxy was born and how it has evolved.
The theoretical overview let us conclude that there are questions which were raised up decades ago and are still unanswered. Other problems were solved over 1-2 decades. In any case the solution of theoretical issues requires a much longer period than the duration of the supports of the funding agencies.
The past shows that new data always inspired 
theoretical improvements. Hopefully, the space data will accelerate work on the theoretical issues. The overview shows that many 
similar physical processes were localized in pulsating stars of different
evolutionary status. Amplitude and/or period variations were found among 
nearly all types of non-stochastically excited variables.
Period doubling was localized in RR Lyraes, Cepheids and white dwarfs.
Mode trapping was used as an explanation in RR Lyrae, white dwarfs and sdB stars.
Non-linear coupling and resonance
were mentioned in different types of stars as a possible explanation for the
observed phenomenon. Outbursts were localized in Be stars and white dwarfs, although the mechanisms are different. 
Maybe the similar phenomenon obtained from stars of different evolutionary status will help us in find the physical cause and let us get closer to the general aspect of pulsation.

Nowadays, we do not expect
to discover major new classes of pulsating variable stars, that would add drastically to our knowledge of pulsation, 
although there are still places for new groups on the HR diagram.
As 
\citet{Kilkenny1997} concluded after the discovery of sdB stars, "serendipity
appears to have a major role to play in research and we are forcibly reminded
that if we only look for what we expect to find, we might well miss
exciting new discoveries."
Nevertheless, the greatest challenge is to
step further in describing the physics of the stars. \citet{CoxJP1976} 
summarized the difficulties connected to the non-radial pulsation of
$\delta$ Scuti stars reminding us that the fully non-linear, non-radial,
non-adiabatic calculation of stellar oscillations has not been attempted by
anyone. The theoretical framework and application in 
simplified situations will be needed to guide physical intuition first. 
The full set of non-linear partial differential equations that describe
the pulsation of stars has not been solved without significant approximations
or special assumptions for any star. Of course, since that time some special
assumptions have been relaxed, but the homework is clear. If we cannot give a reliable description of a real
star, not in average, but from single star to single star, then modeling
is not mature enough to help the interpretation of the observed data. In 
"real" stars we are faced with not only pulsation, but different physical
processes: rotation, diffusion, convection, magnetic field, outbursts, spots,
flares, everything that we see in the case of our Sun. Space data provided
this high level of knowledge.

Computer-minded and/or sensitive thinking young scientists are needed who are
able to process and interpret the flood of space data. The need for computer-minded young scientists is because a larger and larger database is going to be treated and new pulsation codes are expected to  be written to include more and more, up to now excluded, physical processes. In addition, 
new ways of looking at the data and the theory together are required compared to what was used in the past. 

Maybe we cannot immediately give the age of the Universe, but the future of
asteroseismolgy is definitely bright. It depends on us how open-minded we are
to notice previously unpredicted (by theory) and unidentified (by observation)
features of the Universe.

\section*{Funding}
This work was supported by the ESA PECS Grant No 4000103541/11/NL/KML. 
\section*{Acknowledgments}
Many thanks to J.M. Benkő for his support on solution of technical problems. 
This research has made use of NASA's Astrophysics Data System Bibliographic Services. Thanks to the anonym reviewers and especially thanks to Paul Bradley,
who offered a great help during his review process. I am grateful to the co-editor, 
Joyce Guzik, who was interested in my opinion and invited me to write this
paper.

\bibliographystyle{mnras}
\bibliography{paparo}
\end{document}